\documentclass[%
 reprint,
 amsmath,amssymb,
 aps,
]{revtex4-2}
\usepackage{}
\usepackage{hyperref}
\usepackage{graphicx}
\usepackage{dcolumn}
\usepackage{bm}
\usepackage{lineno}
\usepackage{color}
\usepackage{comment}

\begin{document}

\preprint{APS/123-QED}
\title{Superposition model for energy reconstruction and mass identification in cosmic ray spectra
}
\author{Hu Liu$^{1}$}
\email{huliu@swjtu.edu.cn}
\author{Fanping Li$^{1}$}
\author{J. Zhao$^{2,3}$}
\author{L. Y. Wang$^{2,3}$}
\author{Zhe Li$^{2,3}$}
\author{S. Z. Chen$^{2,3}$}
\address{
$^1$ School of Physical Science and Technology, Southwest Jiaotong University, Chengdu 610031, Sichuan, China \\
$^2$ Key Laboratory of Particle Astrophysics \& Experimental Physics Division \& Computing Center, Institute of High Energy Physics, Chinese Academy of Sciences, Beijing 100049, China \\
$^3$ TIANFU Cosmic Ray Research Center, Chengdu, Sichuan, China
}


\begin{abstract}
The ``knee'' of cosmic ray spectra may reflect the maximum energy accelerated by galactic cosmic ray sources or the limit of the galaxy's ability to bind cosmic rays. Measurements of individual energy spectra are a crucial tool to understand the origin of the knee. Energy reconstruction and composition identification are foundations of the individual energy spectra measurements. One of the main scientific goals of Large High Altitude Air Shower Observatory (LHAASO) is measuring the cosmic ray energy spectra and composition from $\sim$ 10 TeV to $\sim$ EeV. In this work, a novel method for reconstructing energy and logarithm mass (lnA) based on a superposition model is introduced. Energy and lnA are reconstructed using two universal, composition- and energy-independent calibration lines. For zenith angle below 40 degree, the energy and lnA biases are within $\pm$5\% and $\pm$0.3, respectively, across all compositions. The method uses particle densities—measured by LHAASO’s electromagnetic and muon detectors at a fixed distance from the shower axis—rather than integrated particle counts in annular bands. The density-based approach improves resolution for both energy and lnA, especially for heavy nuclei. The resulting energy resolution ranges from below 5\% to $\sim 15\%$ above 1 PeV, the best mass resolution for iron achieved is below 25\% above 10 PeV. The hadronic model dependencies of energy and lnA are also reported. These dependencies scale with lg(E/A) and are nearly independent of primary composition.
\end{abstract}

\maketitle

\section{Introduction}
A number of significant features have been observed in the cosmic ray spectrum, most notably the so-called "knee" at around 3 × $10^{15}$ eV and the "ankle" at around 5 × $10^{18}$ eV~\cite{BLUMER2009293,lhaaso-all}. The origin of the knee remains under debate; however, it is widely considered to provide strong constraints on acceleration and propagation models~\cite{Erlykin_1997, Berezhko_2007}. The region between the knee and the ankle is regarded as the transition from Galactic to extra-galactic origins~\cite{Hillas_2005}. Determining the positions of the knees in the spectra of different cosmic-ray species is therefore a crucial tool for investigating their acceleration and propagation mechanisms.\\ \indent
Currently, space and balloon-borne experiments are limited to measuring cosmic rays up to energies of $\sim$ $10^{14}$ eV~\cite{AGUILAR20211, CALETProton, DAMPEProton, DAMPEHelium, ISS-CREAM}, due to constraints on payload size and weight. Cosmic rays in the ``knee'' energy range have been studied exclusively by ground-based experiments, which detect the extensive air showers (EAS) produced by primary particles entering the atmosphere. In these EAS experiments, the power to discriminate between different nuclear species is limited compared to direct measurements by space and balloon-borne experiments~\cite{Composition}. While many ground-based experiments have measured energy spectra by categorizing nuclei into groups (e.g., proton, helium, proton+helium, CNO, MgAlSi, and iron groups)~\cite{lhaaso-proton,TIBET, ARGO, KASCADE, TALE, TUNKA, ICECUBE,KM2A_ind_flux}, contamination from nuclei outside the target group cannot be ignored~\cite{ARGO, KASCADE, TALE, TUNKA, ICECUBE}.In EAS experiments, the measured variables are both correlated with energy and composition of the primary particle (e.g. electron number, muon number, shower maximum, and et al.), that means the energy reconstruction depends on the assumption of the composition of primary particle, the composition sensitive variable depends on the knowledge of energy of primary particle, (e.g. the composition sensitive variable need to be corrected by the reconstructed energy). Most importantly, the reconstruction methods for both energy and composition depend on Monte Carlo (MC) simulations, particularly on the hadronic interaction models used to simulate shower development. These factors collectively limit the precision of energy spectrum measurements for both all-particle and individual cosmic ray species.\\ \indent
Several ground-based methods exist to estimate primary cosmic ray energy. One relies on shower size or electromagnetic secondary counts; however, these quantities weakly depend on composition, introducing up to 20\% systematic uncertainty. Examples include Tibet-III~\cite{TIBET3} and IceTop~\cite{ICETOP}. Another method uses two secondary particle types: electromagnetic (or Cherenkov) counts and muon counts. A novel composite variable such as $N_{e}+25N_{\mu}$ ($N_e$: electromagnetic particle count; $N_{\mu}$: muon count) strongly correlates with primary energy and is nearly composition-independent, yielding energy biases within $\sim5\%$ across all compositions. Representative experiments include CASA-MIA \cite{CASA-MIA} and LHAASO \cite{lhaaso-energy-km2a,lhaaso-energy-wfcta}. Ultra-high energy cosmic ray experiments adopt a calorimetric approach using air fluorescence light to reconstruct primary energy, with corrections for invisible energy. Composition-induced uncertainty is $\sim8\%$ , as demonstrated by the Pierre Auger~\cite{AUGER2} and Telescope Array (TA)~\cite{TA}. Two standard approaches are available for logarithmic primary mass reconstruction. The first leverages the linear correlation between fluorescence-measured shower maximum and log mass at fixed energy~\cite{Composition}; the second employs muon counts~\cite{lhaaso-all}, which also vary linearly with log mass for a fixed energy. Both methods demand pre-reconstructed primary energy. \\\indent
The Large High Altitude Air Shower Observatory (LHAASO), located at Haizishan, Daocheng, Sichuan province, China (4410 m above sea level)~\cite{LHAASO_Design}, consists of 1.3 $km^{2}$ array (KM2A) of electromagnetic particle detectors (EDs) and muon detectors (MDs) for $\gamma$-ray astronomy above 10 TeV and cosmic ray physics, a water Cherenkov detector array (WCDA) with a total active area of 78,000 $m^{2}$ for TeV $\gamma$-ray astronomy, 18 wide field-of-view air Cherenkov/fluorescence telescopes (WFCTA) for cosmic ray physics from 10 TeV to $\sim$EeV energies, and a newly proposed electron-neutron detector array (ENDA) covering 10,000 $m^{2}$ for cosmic ray physics~\cite{LHAASOCpt1}. LHAASO comprehensively measures air showers by detecting several types of secondaries. For cosmic-ray physics, the WCDA provides an absolute energy scale calibration below 35 TeV by observing the Moon shadow deficit and measures composition-sensitive variables up to around 10 PeV~\cite{WCDACal,YouICRC}. The WFCTA measures the longitudinal development of EAS, its energy reconstruction and composition discrimination of nuclei have been studied in reference~\cite{WFCTARec,lhaaso-energy-wfcta,lhaaso-proton,ICRCLong}. KM2A measures the lateral distribution of EAS, energy and $\ln A$ reconstruction using electromagnetic and muon particle counts has been studied previously~\cite{lhaaso-energy-km2a,lhaaso-all}. In this work, we present a novel method for reconstructing energy and $\ln A$ of primary nuclei. This method is based on a superposition model and utilizes electromagnetic and muon particle densities instead of counts. \\\indent
This work is organized as follows: the detector description and simulation parameters are introduced in Section~\ref{experiment}, the reconstruction methods are presented in Section~\ref{evtrec}, the performance of the energy and $\ln A$ measurements is evaluated in Section~\ref{performance}, and finally, a summary is given in Section~\ref{summary}.

\section{Experiment}
\label{experiment}
\subsection{Detector}
KM2A is a surface detector array composed of two kinds of detector array: the electromagnetic particle detector (ED) array and the muon detector (MD) array.The ED array is uniformly distributed over an area of 1.3 $km^{2}$. It is divided into two parts: a central array, consisting of 4911 EDs with 15 m spacing within a circular area of radius 575 m, and a guard-ring array, consisting of 305 EDs with 30 m spacing surrounding the central one, extending to an outer radius of 635 m. The detailed specifications of the ED are provided in reference~\cite{LHAASOCpt1}.  \\ \indent
An ED unit consists of four plastic scintillation tiles (each measuring 100 cm $\times$ 25 cm $\times$ 1 cm). Each tile is covered by a 5-mm-thick lead plate to absorb low-energy charged particles in the shower and to convert $\gamma$-rays into electron-positron pairs, thereby improving the angular and core position resolution of the array. When a high-energy charged particle enters the scintillator, it deposits energy and excites the scintillation medium, producing numerous scintillation photons. These photons are collected by embedded wavelength-shifting (WLS) fibers and transmitted to a 1.5-inch photomultiplier tube (PMT). The PMT records the arrival time and the number of photons. The time resolution of an ED unit is approximately 2 ns. The charge resolution for measuring a single particle is better than 25 \%, and the dynamic range extends from 1 to $10^4$ particles. Further details regarding the ED design can be found in reference~\cite{LHAASOCpt1}.  \\ \indent
The MD array is composed of 1188 water Cherenkov tanks deployed on a grid with a spacing of 30 m. An MD is a pure water Cherenkov detector enclosed within a cylindrical concrete tank with an inner diameter of 6.8 m and height of 1.2 m. An 8-inch PMT is installed at the center of the top of the tank to collect the Cherenkov light produced by high energy particles traversing the water. The whole detector is covered by a steel lid underneath soil. The thickness of the overburden soil is 2.5 m, to absorb the secondary electrons/positrons and $\gamma$-ray in the shower. Thus the particles that can reach the water inside and produce detectable Cherenkov signals are almost exclusively muons, except for those MD located at the very central part of showers where some very high energy EM components may have a chance to punch through the screening soil layer (noted as ``punch-through'' effect below). The charge resolution of the particle counter is \textless 25\% for a single muon and the dynamic range extends from 1 to $10^{4}$ particles. Further details about the MD detector can be found in reference~\cite{LHAASOCpt1}. \\ \indent
The full KM2A array has been operating since July 2021. Its trigger logic has been well tested, more details can be found in reference~\cite{KM2ATrigger}.  \\ \indent

\subsection{Simulation}
\label{simulation}
The simulations performed in this study include the detailed development of air showers in the atmosphere and the response of the KM2A detector. Air showers were generated using the CORSIKA software (v77410)~\cite{Corsika}. For hadronic interactions, the high-energy models QGSJET-II-04, EPOS-LHC, and SIBYLL 2.3d were used in conjunction with FLUKA for low energies, while electromagnetic interactions were handled by the EGS4 model. The plots in this work are based on data simulated with the QGSJET-II-04 model unless otherwise specified.Five components–proton, helium, CNO group (the mass number, noted as A, is 14), MgAlSi group (A is 27), and iron are generated with the following energies and zenith angles. Events with energies from $10^{14}$ to $10^{17}$ eV are simulated. They follow an isotropic angular distribution, with zenith angles of 0$^{o}$–40$^{o}$ and azimuths of 0$^{o}$–360$^{o}$. The shower core positions were uniformly distributed within a circle of 1000 m radius, centered on the KM2A array. The height of observing ground is 4410m.
The secondary particles reaching ground level were processed by a dedicated detector simulation program (named G4KM2A~\cite{KM2ASim, KM2ASim2,KM2ASim3}) in the framework of Geant4 package~\cite{Geant4}. \\\indent 
The reliability of the detector simulation was verified via the KM2A array data~\cite{KM2ASim2,KM2ASim3}, detailed comparison between data and MC simulation is beyond the scope of this paper.
\section{Reconstruction}
\label{evtrec}
The reconstruction of KM2A events is similar to the procedures described in reference~\cite{KM2ACrab}. Each shower event is composed of many ED and MD hits, each of which has timing and charge information. In combination with the positions of these detectors, the primary direction, core location and lateral distribution of the shower event can be reconstructed. For KM2A events, The ED hits are used for direction and core location reconstruction. The lateral distribution reconstructed from ED hits and MD hits is used for energy and lnA reconstruction. To combine all the information provided by ED and MD, two formulas were used to fit the lateral distribution of ED and MD respectively during the reconstruction, details about the reconstruction are listed below.\\ \indent
\subsection{Lateral distribution}
\label{rec_method}
\begin{figure*}[htbp]
\begin{minipage}[t]{1.\linewidth}
\includegraphics[width=0.45\linewidth]{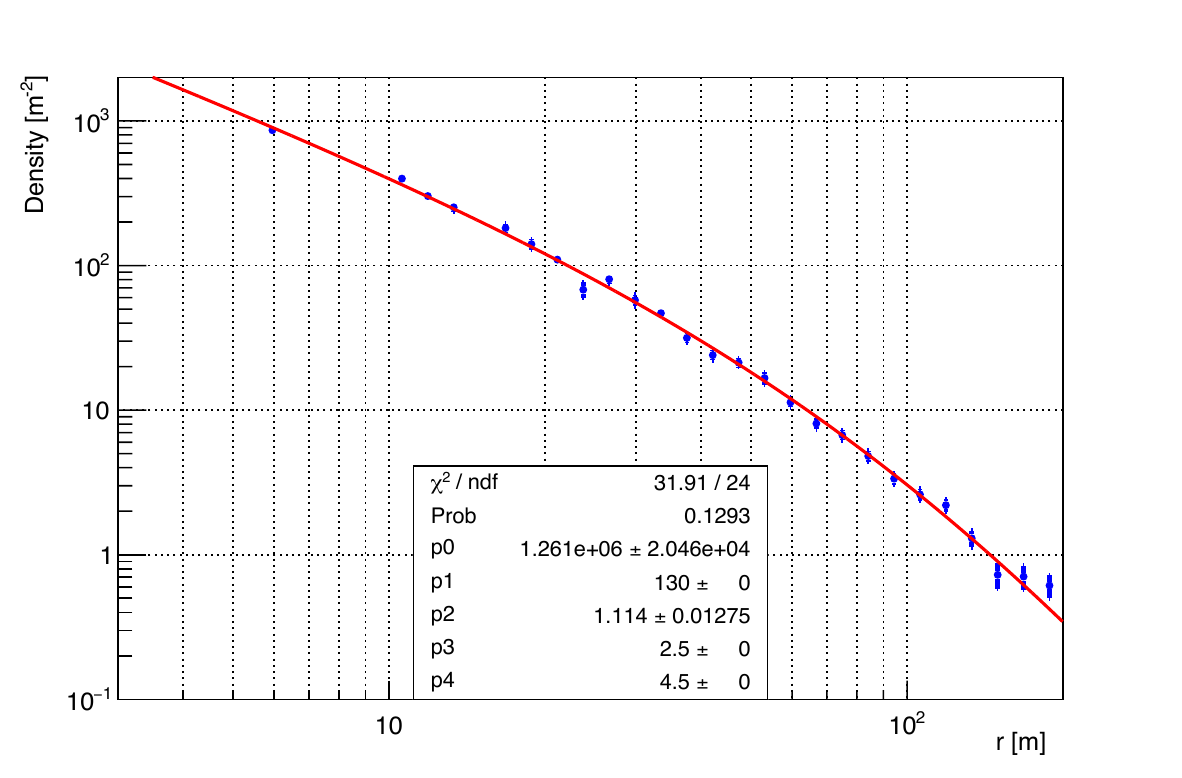}
\hspace{0cm}
\includegraphics[width=0.45\linewidth]{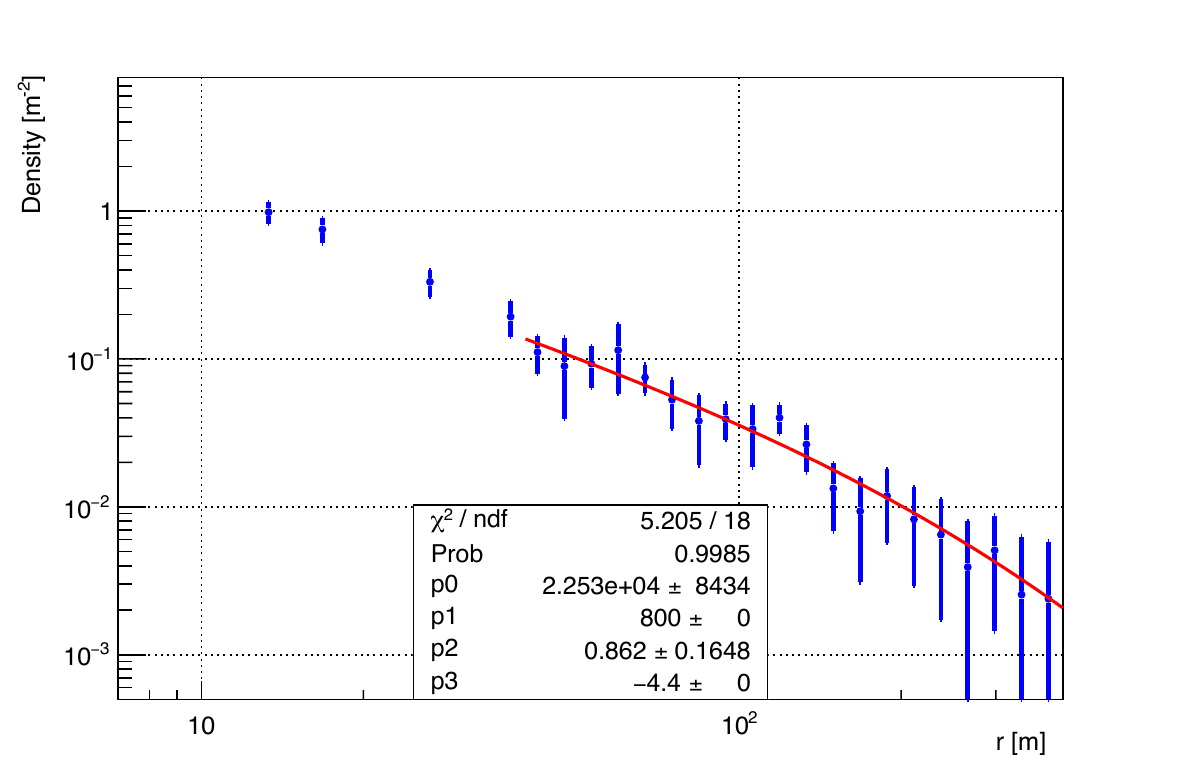}
\caption{Lateral distribution (density vs r, r is the perpendicular distance to shower axis) reconstructed from ED (left) and MD (right), the primary particle is proton with energy equal to 1.3 PeV, the fitting result is also shown as the red line. The fitting formula for ED is equation \ref{eq1}, and the fitting formula for MD is equation \ref{eq2}.}
\label{later_rec}
\end{minipage}
\end{figure*}

For the ED reconstruction, the procedures are the same as those used for $\gamma$-ray, which is described in detail in reference~\cite{KM2ACrab}. First, a time window of 400 ns and a circular window with a radius of 100 m are adopted to select the most probable real secondary shower hits. With these selected hits, the core location is reconstructed using an optimized centroid method and the direction is reconstructed by fitting the shower plane. Second, only hits within a time window of [-30, 50] ns  relative to the shower front and with a distance less than 200 m from the shower core are selected. The equation \ref{eq1} (the same as that used for $\gamma$-ray in reference~\cite{KM2ACrab}) is used to fit the ED lateral distribution, where $\rho_{em}$ is the density of particles measured by the ED, $r_m$ is the Moliere radius, it was fixed at 130 m, and r is the perpendicular distance to the shower axis, Using these selected hits, the core location, the number of particles (i.e., shower size, $N_{e}^{size}$ in equation \ref{eq1}), and shower age (s in equation \ref{eq1}) are reconstructed via a maximum likelihood method, the direction is also updated using the new core location. Several iterations are applied to improve the reconstruction performance. Fig. \ref{later_rec} shows the lateral distribution of ED and the corresponding fitting result.
\begin{equation}
    \rho_{em}(r) = \frac{N_{e}^{size}}{2\pi r_{m}^{2}} \frac{\Gamma(4.5-s)}{\Gamma(s-0.5) \Gamma(5-2s)} (\frac{r}{r_m})^{s-2.5}(1+\frac{r}{r_m})^{s-4.5} \tag{1}
    \label{eq1}
\end{equation}
\indent
Following the ED reconstruction, the MD lateral distribution is reconstructed using the MD hits, combined with the primary direction and core location obtained from the ED reconstruction. For MD, only hits within a time window of [-30, 50] ns relative to the shower plane are selected, to reduce the punch-through effect from high energy electromagnetic secondaries, only hits located between 40 m and 400 m from the shower axis are selected. The MD lateral distribution is fitted with equation \ref{eq2} using a maximum likelihood method similar to that employed for the ED reconstruction. Equation \ref{eq2} is similar in form to the one used for ED reconstruction, it was derived from CORSIKA data, details can be found in reference~\cite{EASliu,EASliu2}. Where $\rho_{\mu}$ is the density of muons measured by MD, $r_{0}$ is a constant fixed at 800 m based on the muon lateral distribution from CORSIKA data~\cite{EASliu,EASliu2}, $C_{\mu}$ is the normalize parameter, $s_{\mu}$ describes the shape of lateral distribution. $C_{\mu}$ and $s_{\mu}$ are the two free parameters determined from the fit. The muon shower size (denoted as $N_{\mu}^{size}$ below) can be obtained by integrating the density over the entire area. Fig. \ref{later_rec} (right) shows the lateral distribution of MD and the corresponding fitting result. 
\begin{equation}
    \rho_{\mu}(r) = \frac{C_{\mu}}{2\pi r_{0}^{2}} (\frac{r}{r_0})^{s_{\mu}-2}(1+\frac{r}{r_0})^{-4.4} \tag{2}
    \label{eq2}
\end{equation}
\\\indent

\subsection{Variables for energy and lnA reconstruction}
\begin{figure*}[htbp]
\begin{minipage}[t]{1.\linewidth}
\includegraphics[width=0.45\linewidth]{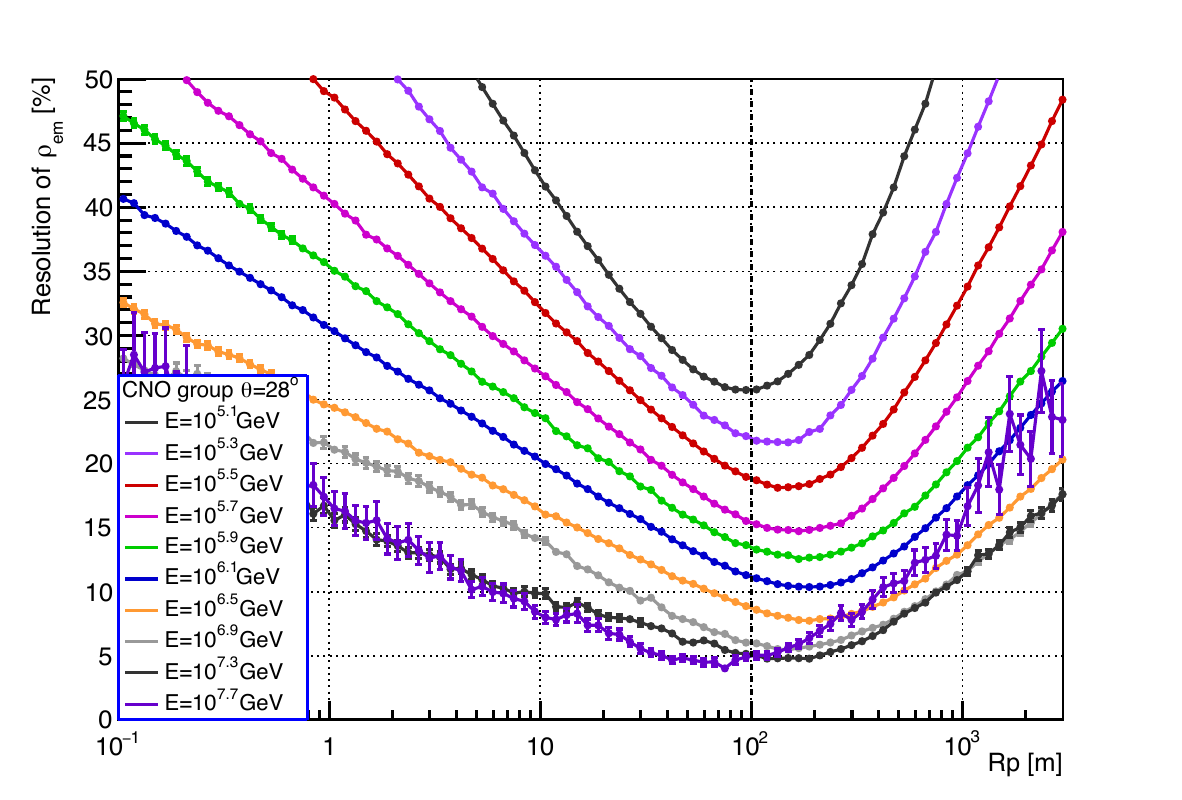}
\hspace{0cm}
\includegraphics[width=0.45\linewidth]{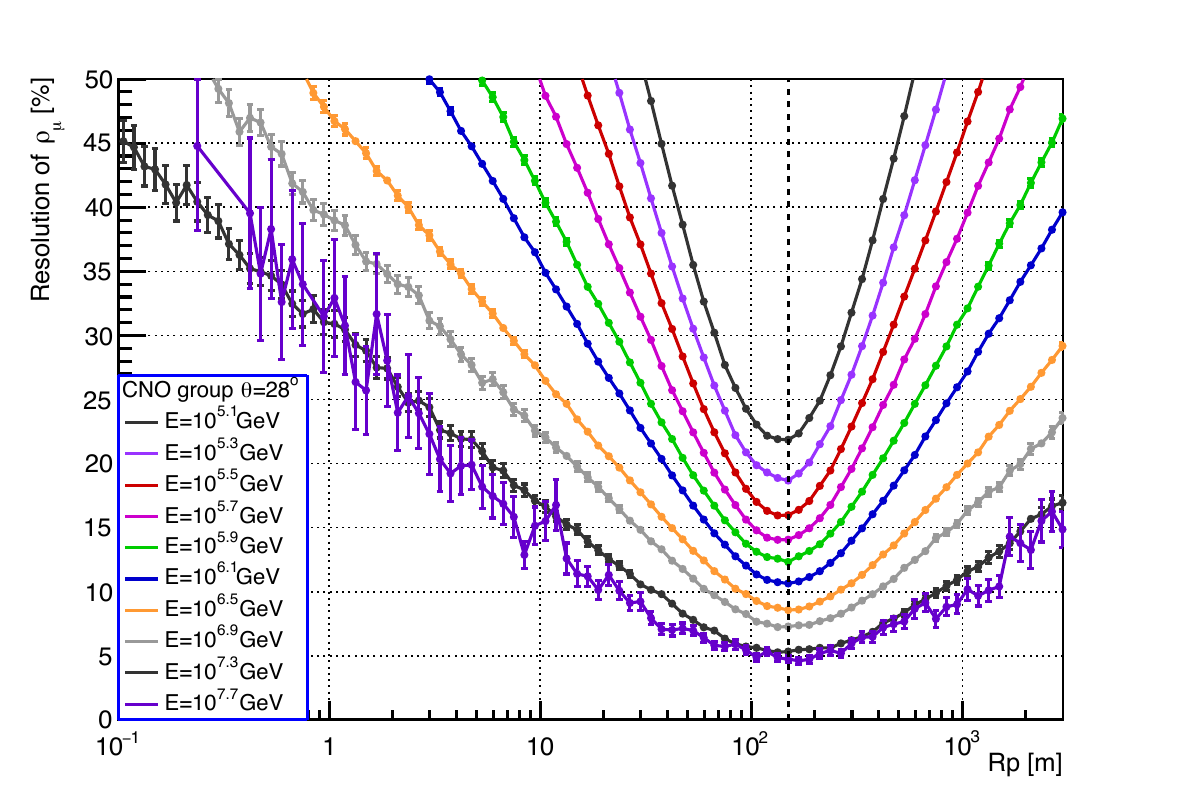}
\caption{The resolution of density from ED ($\rho_{em}$) calculated from equation \ref{eq1} (left) and density from MD ($\rho_{\mu}$) calculated from equation \ref{eq2} (right) vs r (the perpendicular distance to the shower axis) for a CNO sample with zenith angle $\theta=28^{o}$. The resolution of density is defined as the sigma of fitted Gaussian function to distribution of logarithm density. Different lines correspond to different energy indicated in the legend of plot, the dashed vertical line indicates the chosen distance that have the best resolution of density, which is 100 m for ED and 150 m for MD, the results are similar for other primary particles and other zenith angles.}
\label{size_rec}
\end{minipage}
\end{figure*}
\begin{figure*}[htbp]
\begin{minipage}[t]{1.\linewidth}
\includegraphics[width=0.45\linewidth]{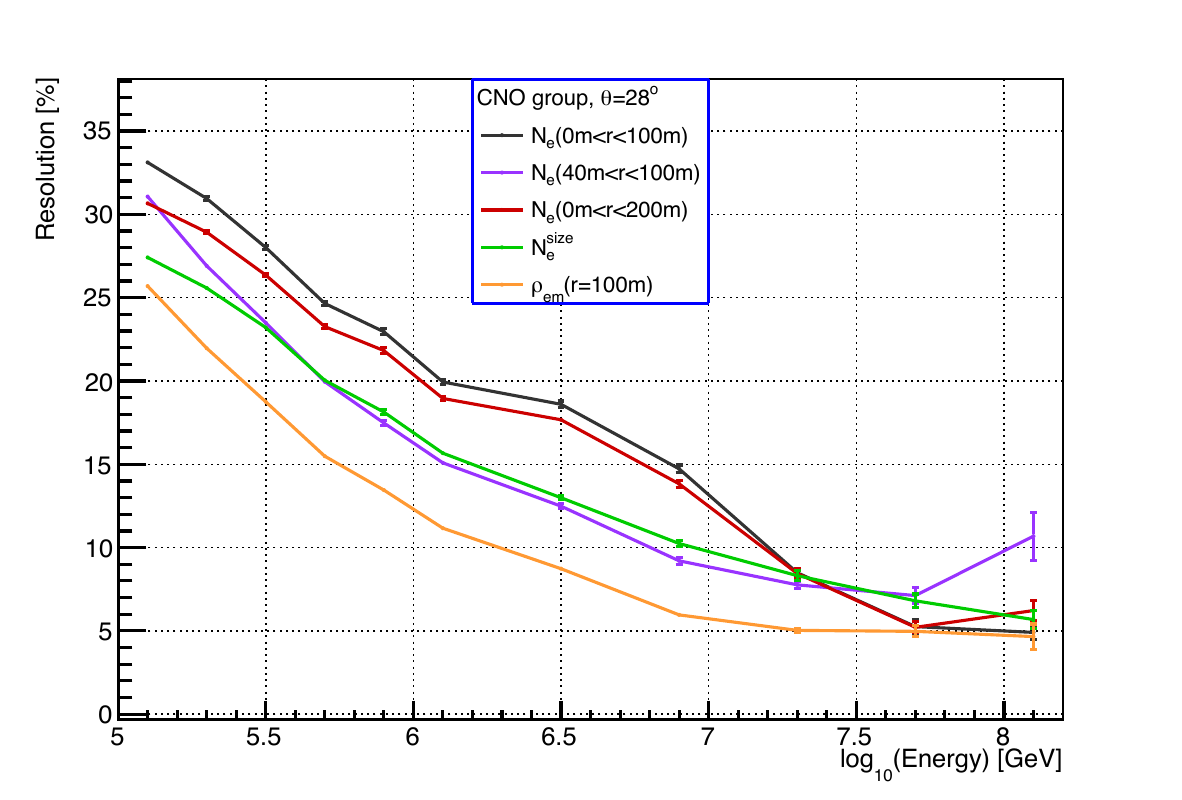}
\hspace{0cm}
\includegraphics[width=0.45\linewidth]{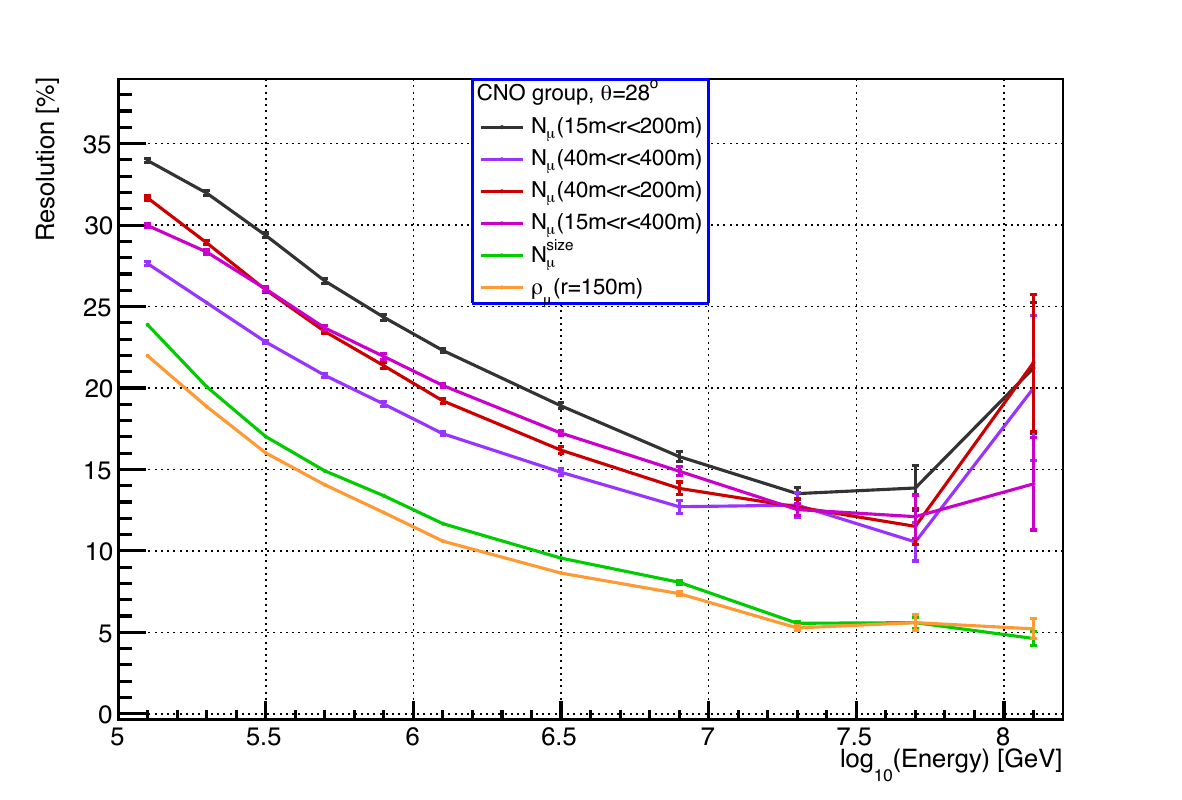}
\caption{The resolution comparison of several variables for ED (left) and MD (right) as a function of the true energy. The orange lines are the density at 100 m for ED (left) and density at 150 m for MD (right) respectively, the green lines are the shower size (the integration of lateral distribution function over area) of the fitted lateral function, all the other lines are the number of particles detected by ED and MD in an annular band indicated in the legend of the plot. The primary particle is CNO group, and the zenith angle is $\theta$=28$^{o}$, the results are similar for other primary particles and other zenith angles.}
\label{res_vars}
\end{minipage}
\end{figure*}

 \indent
By fitting the lateral distribution, all information is combined into two variables, $N_{e}^{size}$ and s for ED, $C_{\mu}$(or $N_{\mu}^{size}$) and $s_{\mu}$ for MD. $N_{e}^{size}$ (or $N_{\mu}^{size}$) has additional fluctuation due to the number of secondary electrons (or muons) depends on the stage of shower development. However, the shape of lateral distribution also depends on the stage of shower development~\cite{EASliu}. At some distance to the shower axis, these two effects may compensate for each other, and the shower to shower fluctuation at this distance reaches a minimum. In fact, as demonstrated by several experiments~\cite{ICECUBE,ASgamma_density,HAWC_density,AUGER}, the particle density at a fixed distance from the shower axis serves as an excellent estimator of the primary energy. Fig. \ref{size_rec} shows the resolution of electromagnetic density from ED ($\rho_{em}$, calculated from equation \ref{eq1}) and the  muon density from MD ($\rho_{\mu}$, calculated from equation \ref{eq2}) versus the perpendicular distance to the shower axis for a sample of CNO group events at several energies, the resolution of density is defined as the sigma of a Gaussian function fitted to the distribution of the logarithm of the density, to remove the fluctuation induced by the bin width of energy and zenith angle of primary particle, the zenith angle is fixed to be $28^{o}$, and energy of the sample is fixed to a value shown in the legend of Fig. \ref{size_rec}. As can be seen, the distances that yield the best resolution (or minimum fluctuation) for $\rho_{em}$ and $\rho_{\mu}$ are around 100 m and 150 m respectively. The distance for other primary particles and other zenith angles are also similar. To simplify the reconstruction, 100 m for density measured by ED (denoted as $\rho_{ED}$ below) and 150 m for density measured by MD (denoted as $\rho_{MD}$ below) were chosen for all types of primary particles in all energy range and zenith angle range in this work. \\\indent
There are also many other variables that were widely used for energy reconstruction and composition identification in many experiments, e.g., the number of particles detected in an annular band and the shower size~\cite{KM2ACrab}. Fig. \ref{res_vars} shows the comparison of the resolution or fluctuation (defined as the sigma of a Gaussian function fitted to the logarithm distribution) between those variables and the density at the distance derived from Fig. \ref{size_rec}, for a sample of the CNO group, to remove the fluctuation induced by the bin width of energy and zenith angle of the primary particle, the zenith angle is fixed at $28^{o}$, and the x-axis represents the fixed energy value. As can be seen, $\rho_{em}$ at r=100 m has the best resolution compared to all other variables, $\rho_{\mu}$ at r=150 m is slightly better than the muon shower size and is much better than the other variables. This is also true for other primary particles and zenith angles. Therefore, the two variables of $\rho_{ED}$ and $\rho_{MD}$, derived from the fitted lateral distributions of ED and MD, will be used for energy and lnA reconstruction, since the energy and lnA resolution are both directly related to the resolution of the variables used.\\

\indent

\begin{figure*}[htbp]
\begin{minipage}[t]{1.\linewidth}
\includegraphics[width=0.9\linewidth]{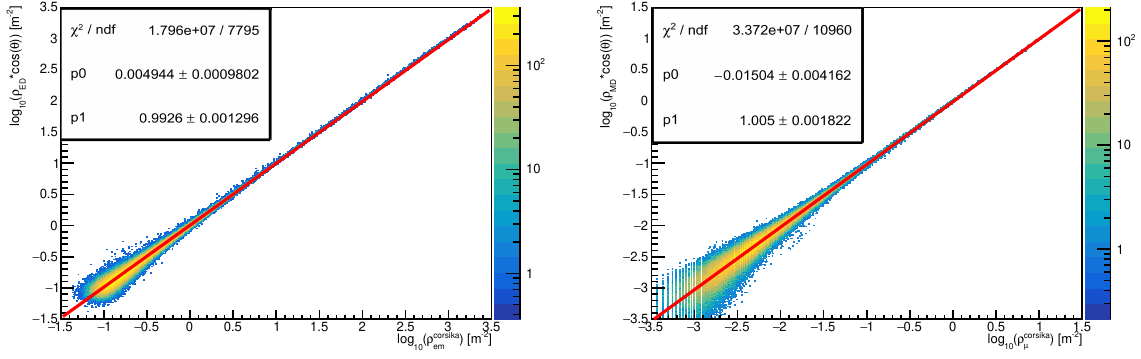}
\hspace{0cm}
\caption{The relationship between the reconstructed density for ED(left) and MD(right) from detector responded data (y axis) and the density counted from CORSIKA data (x axis) for the mixed-composition sample (normalized by H3a flux model~\cite{Gaisser_Model}). The $cos(\theta)$ in y axis is used for correcting the path-length effect in the detector. The red lines are a linear fit to data.}
\label{recsize_corsika}
\end{minipage}
\end{figure*}

To verify the reconstruction method, the comparison of $\rho_{ED}$(or $\rho_{MD}$) multiplied by cos($\theta$) and the electromagnetic number density(or muon density) at the same distance to the shower axis counted in CORSIKA data (denoted as $\rho_{em}^{corsika}$ and $\rho_{\mu}^{corsika}$ below) is presented in Fig. \ref{recsize_corsika} for the mixed-composition sample, the events are normalized by H3a flux model~\cite{Gaisser_Model}, where $\theta$ is the zenith angle, cos($\theta$) accounts for the path-length effect of the detector. 
For this calculation, the threshold energy is set to 10 MeV for secondary electromagnetic particles in CORSIKA. For secondary muons, the threshold energy is set to 1 GeV in CORSIKA, as suggested in reference~\cite{LHAASOCpt1}. The reconstructed $\rho_{ED}$ and $\rho_{MD}$ show a linear relationship with the CORSIKA-derived values, the red line is a linear fit to data. The difference between the reconstructed and true number densities is within $\sim$5\% both for electromagnetic particles and muons. \\\indent

\subsection{Superposition model}
The superposition model is a simplified approach for describing the interaction of a composite nucleus (like iron or oxygen) with a target. It approximates the nucleus as a collection of independent nucleons (protons and neutrons), each carrying a fraction of the total energy. The resulting air shower is then modeled as the superposition of many individual nucleon-induced sub-showers, effectively scaling the profile of a proton-initiated shower of lower energy. According to this model, the number of secondaries can be expressed as in Equation \ref{superpos}, where $N_{e/\mu}^{A,E}$ is the number of electrons (or muons) produced by primary particle mass number A and primary energy E (the base-10 logarithms of the mass number and energy are denoted $lgA$ and $lgE$ hereafter), $N_{e/\mu}^{P,E/A}$ is the number of electrons (or muons) produced by primary proton and primary energy $E/A$.
\begin{equation}
    N_{e/\mu}^{A,E}=AN_{e/\mu}^{P,E/A} \tag{3}
    \label{superpos}
\end{equation}
\begin{equation}
    lg(N_{e/\mu}^{A,E}/A)\sim lg(E/A)=lg(N_{e/\mu}^{P,E/A})\sim lg(E/A) \tag{4}
    \label{superpos2}
\end{equation}
According to Equation \ref{superpos}, the relationship between lg($N_{e/\mu}^{A,E}/A$) and lg($E/A$) for a primary nucleus of mass number A reduces to lg($N_{e/\mu}^{P,E/A}$) versus lg($E/A$) (equation \ref{superpos2}) (where lg denotes the base-10 logarithm). This represents the number of electrons (or muons) produced by a proton shower versus its primary energy. Consequently, if the superposition model holds, for all compositions, lg($N_{e/\mu}^{A,E}/A$) versus lg($E/A$) follows a single, universal curve: that of a proton. \\\indent
\begin{figure*}[htbp]
\includegraphics[width=0.9\linewidth]{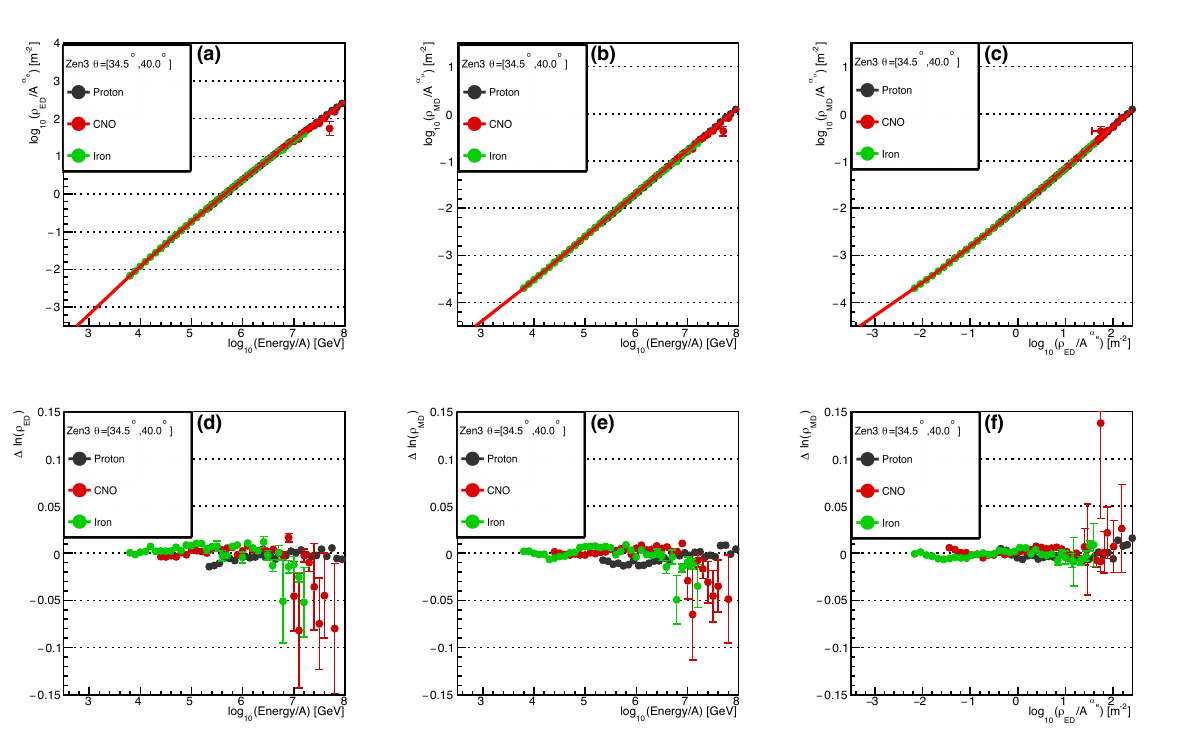}
\caption{The mean relationship between $lg(\rho_{ED}/A^{\alpha_{e}})$ and $lg(E/A)$ (panel a), mean relationship between $lg(\rho_{MD}/A^{\alpha_{\mu}})$ and $lg(E/A)$ (panel b), and mean relationship between $lg(\rho_{MD}/A^{\alpha_{\mu}})$ and $lg(\rho_{ED}/A^{\alpha_{e}})$ (panel c). Panels (d), (e), and (f) present the corresponding residuals.}
\label{Eeu_spos}
\end{figure*}
\begin{equation}
    lg(\rho_{ED})-\alpha_{e}lgA \equiv f_{e}(E/A) \tag{5}
    \label{funED}
\end{equation}
\begin{equation}
    lg(\rho_{MD})-\alpha_{\mu}lgA \equiv f_{\mu}(E/A) \tag{6}
    \label{funMD}
\end{equation}
Fig. \ref{Eeu_spos} shows, for the true mass number A and primary energy E, the relationships derived from MC data of: lg($\rho_{ED}/A$) versus lg($E/A$) (left), lg($\rho_{MD}/A$) versus lg($E/A$) (middle), and lg($\rho_{MD}/A$) versus lg($\rho_{ED}/A$)(right). The upper panels display data points (colored by primary composition) representing mean values per energy bin, with the red curve showing a fit described later. The lower panel shows the deviation of the data points from the fitting curve, or the residual. The relations lg($\rho_{ED}/A$) versus lg($E/A$) for electromagnetic particles and lg($\rho_{MD}/A$) versus lg($E/A$) for muons approximately follow the same curve for all compositions. To align them, small factors $\alpha_{e}$ and $\alpha_{\mu}$ are applied to the mass number A to minimize compositional differences (Eqs. \ref{funED}, \ref{funMD}). As shown in Fig. \ref{Eeu_spos}, the resulting residuals are within about 2$\%$. Both $\alpha_{e}$ and $\alpha_{\mu}$ are close to unity (as expected when the superposition model holds), with $\alpha_{e}$ ranging from 0.99 to 1.01 and $\alpha_{\mu}$ from 1.00 to 1.03, depending on zenith angle. 
Notably, the parameters $\alpha_{e}$ and $\alpha_{\mu}$ are effectively consistent across all Monte Carlo data simulated with different hadronic models (Fig. \ref{EAdiff}), with all compositions following the same curve under these values. We define the universal curves for lg($\rho_{ED}/A^{\alpha_{e}}$) and lg($\rho_{MD}/A^{\alpha_{\mu}}$) versus lg($E/A$) as the functions $f_{e}(E/A)$ and $f_{\mu}(E/A)$ (Eqs. \ref{funED} and \ref{funMD}).\\\indent
Combining the relations lg($\rho_{ED}/A^{\alpha_{e}}$) versus lg($E/A$) and lg($\rho_{MD}/A^{\alpha_{\mu}}$) versus lg($E/A$) shows that lg($\rho_{MD}/A^{\alpha_{\mu}}$) versus lg($\rho_{ED}/A^{\alpha_{e}}$) also follows a universal curve for all compositions, with a residual within about 2$\%$, as shown in fig. \ref{Eeu_spos}c and \ref{Eeu_spos}f.\\\indent
A spline function (shown as the red curve) is used to fit the data points in the upper panels of Fig. \ref{Eeu_spos}. Splines are piecewise functions composed of connected cubic polynomials defined at points called knots, and are widely used in particle physics. A key feature is their smoothness, as they are continuous and have continuous derivatives at the knots. In this work, knots are placed uniformly with the minimum number required, ensuring no clear residual structure. Four knots are used in this work, resulting in five free parameters. \\\indent

\subsection{Energy and lnA reconstruction}
\begin{figure*}[htbp]
\includegraphics[width=0.49\linewidth]{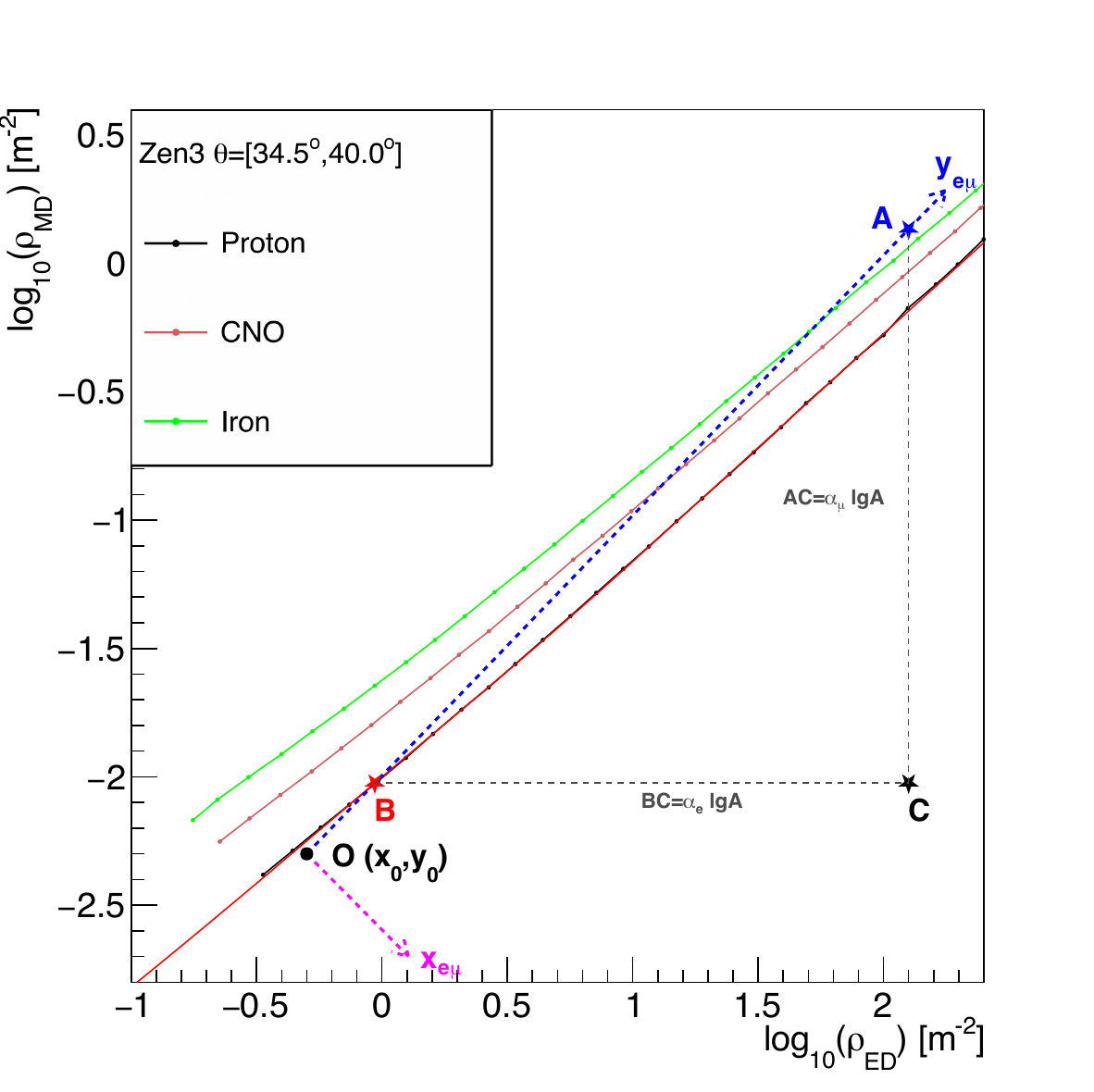}
\caption{Principle of energy and lnA reconstruction. The data points are the mean $lg(\rho_{ED})$ and $lg(\rho_{MD})$ for proton (black), CNO group (red) and iron sample (green). Point A marks the measured values for a single event. The blue dashed line through A is the $y_{e\mu}$ axis (Equation \ref{yeu}); the perpendicular magenta dashed line is the $x_{e\mu}$ axis (Equation \ref{xeu}), with origin at point O. The red curve is from Fig. \ref{Eeu_spos}c. Shifting point A by ($\alpha_{e}lgA$, $\alpha_{\mu}lgA$) aligns it with the curve at point B according to superposition model; their $y_{e\mu}$ difference reconstructs lnA. Since $E/A$ is constant along the $y_{e\mu}$ axis, energy is also reconstructed (see text for details).}
\label{xyeu}
\end{figure*}
Fig. \ref{xyeu} shows the relationship between lg($\rho_{MD}$) and lg($\rho_{ED}$) for different compositions, data points represent mean values per energy bin. Due to shower-to-shower fluctuations, the measured lg($\rho_{ED}$) and lg($\rho_{MD}$) for a single event may not lie on these lines. For example, point A corresponds to such an event. As shown in fig. \ref{Eeu_spos}c, lg($\rho_{MD}/A^{\alpha_{\mu}}$) versus lg($\rho_{ED}/A^{\alpha_{e}}$) follows the proton's curve for all compositions. By extrapolating this phenomenon to all events, point A in fig. \ref{xyeu} should coincide with the proton curve (point B) when shifted by $\alpha_{e}$$lgA$ along the x-axis and by $\alpha_{\mu}$$lgA$ along the y-axis, as indicated by distances BC and AC, respectively. Thus, lnA (where lnA = $lgA$ × ln(10)) can be reconstructed from the distance between points A and B. \\\indent
\begin{equation}
    y_{e\mu} \equiv \frac{\alpha_{e}(lg(\rho_{ED})-x_{0})+\alpha_{\mu}(lg(\rho_{MD})-y_{0})}{\alpha_{e}^2+\alpha_{\mu}^{2}} \tag{7}
    \label{yeu}
\end{equation}
\begin{equation}
    x_{e\mu} \equiv \frac{\alpha_{\mu}(lg(\rho_{ED})-x_{0})-\alpha_{e}(lg(\rho_{MD})-y_{0})}{\alpha_{e}^2+\alpha_{\mu}^2} \tag{8}
    \label{xeu}
\end{equation}
With $y_{e\mu}$ and $x_{e\mu}$ defined as linear combinations of Eqs. \ref{funED} and \ref{funMD} (see Eqs. \ref{yeu} and \ref{xeu}), they are mutually perpendicular and intersect at the origin O. In Fig. \ref{xyeu}, the blue and magenta dashed lines form the corresponding axes, with $x_{0}$ and $y_{0}$ as the coordinates of O. Within the $y_{e\mu}-x_{e\mu}$ coordinates, points A and B differ along the y-axis by $lgA$. Consequently, the quantity $y_{e\mu}-lgA$ is a constant for all events along the $y_{e\mu}$ axis (equal to the y-coordinate of point B) and depends only on $x_{e\mu}$. We define this relationship by Eq. \ref{Arec}, with $f_{A}(x_{e\mu})$ to be determined from MC data. \\\indent
\begin{equation}
    y_{e\mu}-lgA \equiv f_{A}(x_{e\mu}) \tag{9}
    \label{Arec}
\end{equation}
For all events along the $y_{e\mu}$ axis in Fig. \ref{xyeu}, the quantities $lg(\rho_{ED})-\alpha_{e}lgA$ and $lg(\rho_{MD})-\alpha_{\mu}lgA$) are constant, corresponding to the coordinates of point B. Consequently, $E/A$ is also constant (according to Figs. \ref{Eeu_spos}a,\ref{Eeu_spos}b and Eqs. \ref{funED}, \ref{funMD}), implying $y_{e\mu}-lgE$ is constant as well along the $y_{e\mu}$ axis. This constant depends only on $x_{e\mu}$, as defined by Eq. \ref{Erec}, where $f_{E}(x_{e\mu})$ will be determined from MC data. The relationships given by Eqs. \ref{Arec} and \ref{Erec} enable the reconstruction of energy and lnA as follows: \\\indent
\begin{equation}
    y_{e\mu}-lgE \equiv f_{E}(x_{e\mu}) \tag{10}
    \label{Erec}
\end{equation}
\begin{figure*}[htbp]
\includegraphics[width=0.49\linewidth]{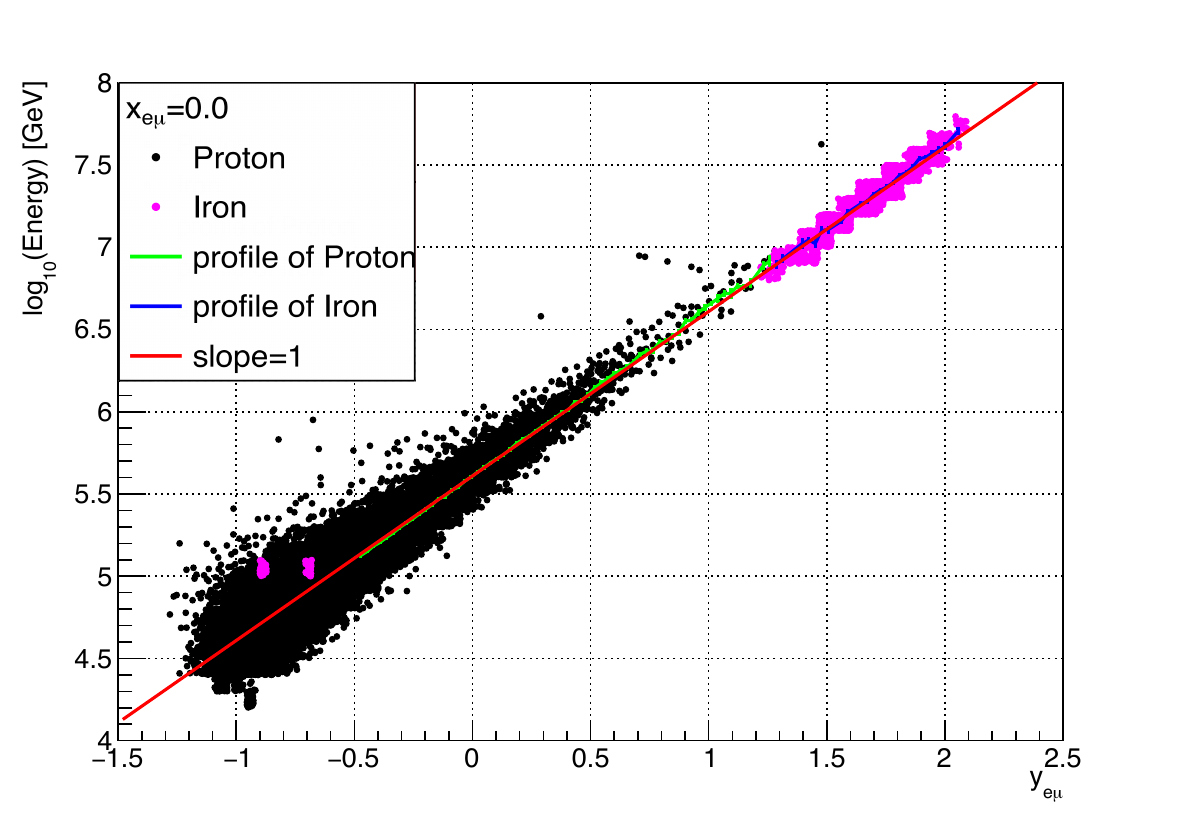}
\includegraphics[width=0.49\linewidth]{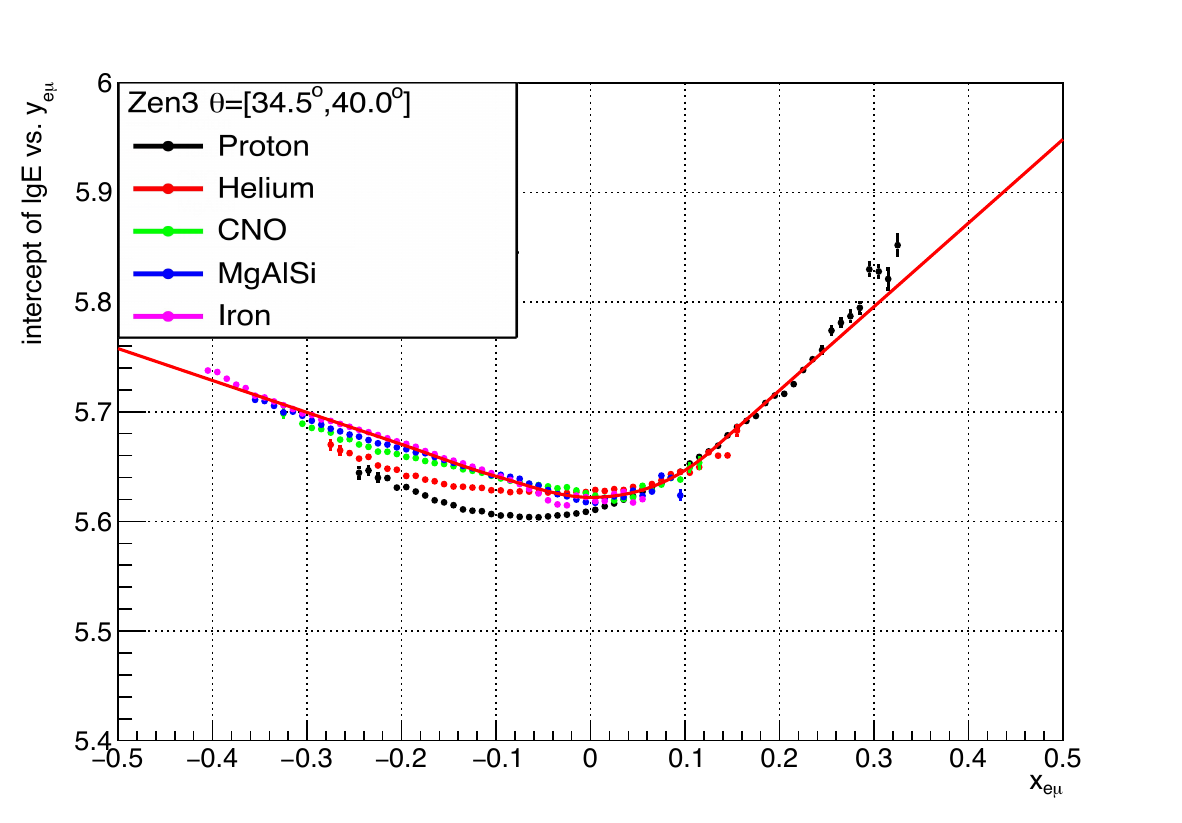}
\caption{Left panel: Primary energy versus $y_{e\mu}$ for $x_{e\mu} \approx 0$. Black and magenta points denote proton and iron samples, respectively; green and blue lines are their corresponding profiles. The red line is a linear fit constrained to unit slope; Right panel: The intercept of the linear fit in left panel as a function of $x_{e\mu}$. Data points are grouped by composition, with a four‑knot spline (red line) fitted to the data.}
\label{lgE_rec}
\end{figure*}
\\\indent
The relationship between $y_{e\mu}$ and $lgE$ in one $x_{e\mu}$ bin is shown in the left panel of fig. \ref{lgE_rec}. Black and magenta points represent the proton and iron samples, respectively, with their profiles shown as green and blue lines. The red straight line is the line with a slope of 1, as predicted by the superposition model. As seen, $y_{e\mu}$ scales linearly with lgE (slope equal to 1), as predicted by Eq.~\ref{Erec}. For each $x_{e\mu}$ bin, the intercept is plotted against $x_{e\mu}$ for different composition in the right panel of fig. \ref{lgE_rec}. They are nearly composition independent and fitted with a spline function with four knots (red line). The red line corresponds to the function $-f_{E}(x_{e\mu})$ from Equation \ref{Erec}, which defines the energy contour line for one energy bin in the $y_{e\mu}$ vs. $x_{e\mu}$ coordinate system. It is universal and independent of primary energy and composition, the main sources of systematic uncertainty in energy reconstruction, defined as the energy reconstruction curve hereafter. \\\indent
\begin{figure*}[htbp]
\includegraphics[width=0.7\linewidth]{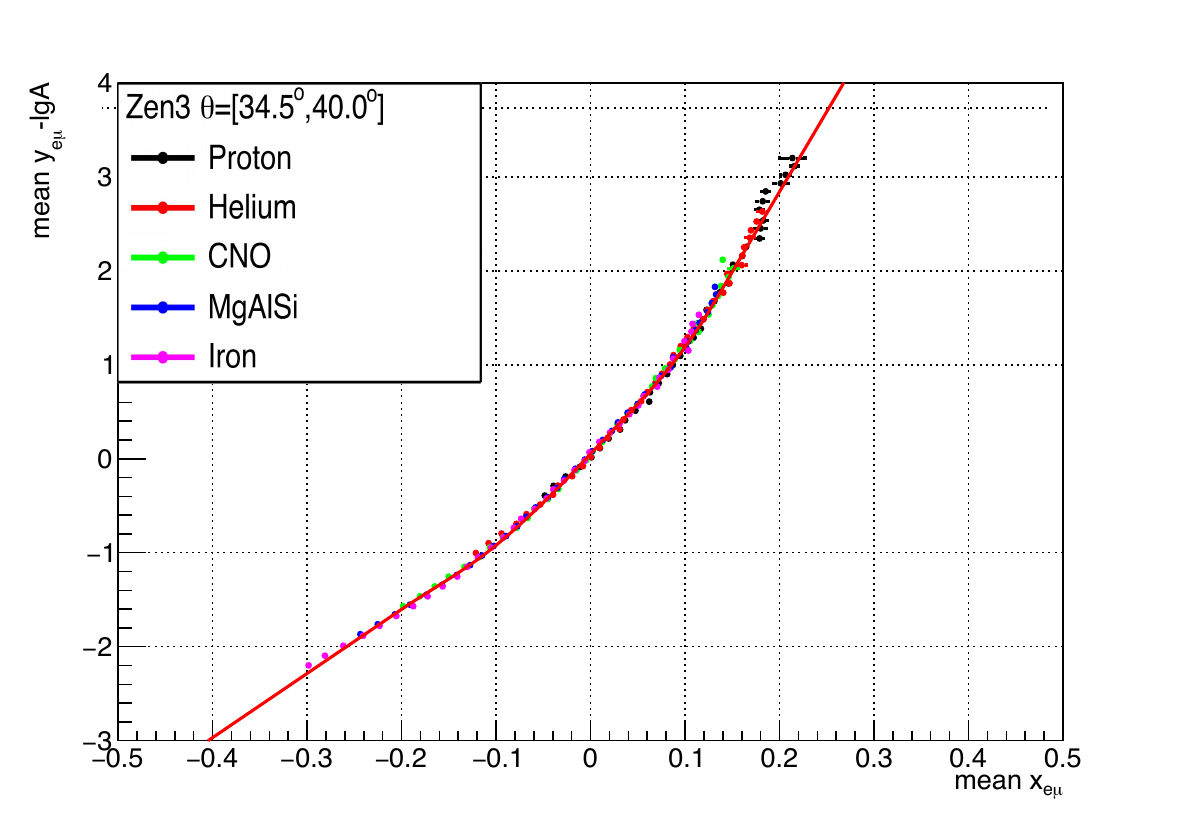}
\caption{Mean $y_{e\mu}-lgA$ versus mean $x_{e\mu}$. Each point is the mean in one reconstructed energy bin, different colors represent different composition. The red line is a spline function fit to data.}
\label{lnA_rec}
\end{figure*}
Fig. \ref{lnA_rec} shows the mean $y_{e\mu}-lgA$ versus mean $x_{e\mu}$, with each point representing the average per reconstructed energy bin, different colors represent different composition. As predicted by Eq. \ref{Arec}, all data follow the same curve regardless of composition. The red curve is a four-knot spline fit to the data, universal and independent of primary energy and composition, the main sources of systematic uncertainty in lnA reconstruction. It represents $f_{A}(x_{e\mu})$ from Eq. \ref{Arec} and is hereafter referred to as the lnA reconstruction curve. \\\indent
\subsection{Hadronic model dependence of electron-muon density}
Energy and lnA reconstruction uses two universal curves, derived from the relationships between $lg(\rho_{MD}/A^{\alpha_{\mu}})$, $lg(\rho_{ED}/A^{\alpha_{e}})$, and $lg(E/A)$ in fig. \ref{Eeu_spos}. These curves depend on the hadronic model used in simulation. Quantifying these differences and their impact on the reconstructed parameters is therefore essential. For simplicity, results from the QGSJet-II-04 model serve as the reference. \\\indent
\begin{figure*}[htbp]
\includegraphics[width=0.9\linewidth]{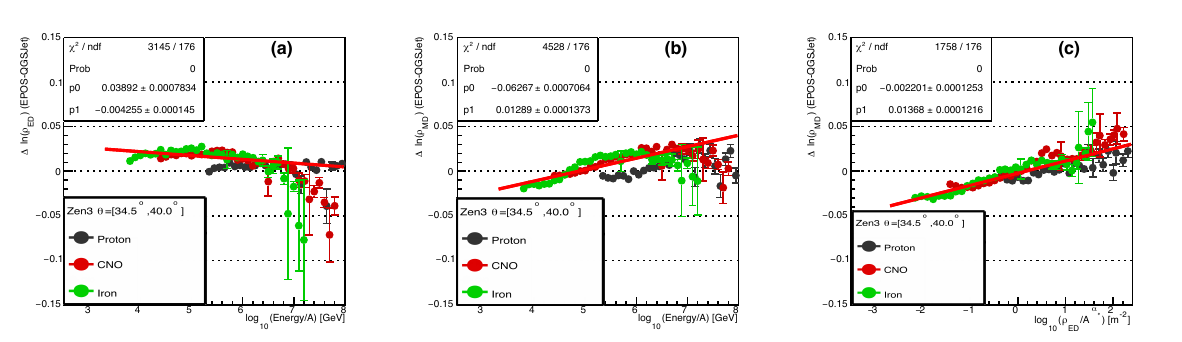}
\hspace{0cm}
\includegraphics[width=0.9\linewidth]{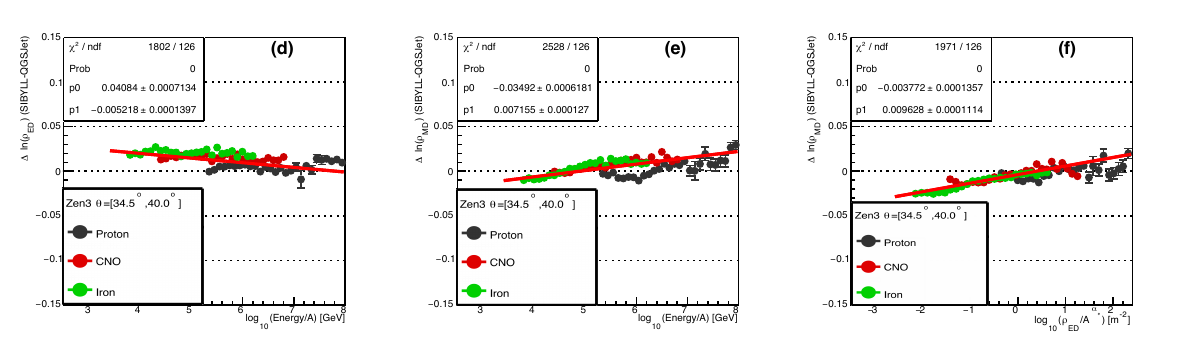}
\caption{Panels (a)–(c): Residuals of the EPOS model relative to QGSJet for (a) $ln(\rho_{ED})$ vs $lg(E/A)$, (b) $ln(\rho_{MD})$ vs $lg(E/A)$, and (c) $ln(\rho_{MD})$ vs $lg(\rho_{ED}/A^{\alpha_{e}})$. Red lines show linear fits to the data. Panels (d)–(f): Residuals of the SIBYLL model relative to QGSJet for (d) $ln(\rho_{ED})$ vs $lg(E/A)$, (e) $ln(\rho_{MD})$ vs $lg(E/A)$, and (f) $ln(\rho_{MD})$ vs $lg(\rho_{ED}/A^{\alpha_{e}})$. Red lines show linear fits to the data. }
\label{EAdiff}
\end{figure*}
Fig. \ref{EAdiff} shows the residuals of the EPOS (panels a–c) and SIBYLL (panels d–f) model data relative to the QGSJet data. For both EPOS and SIBYLL, the residuals show a nearly linear correlation with $lg(E/A)$. The difference between these two models is small, and their deviation from the QGSJet data is within $\sim$ 3\%. Since $lg(\rho_{MD}/A^{\alpha_{\mu}})$ vs. $lg(\rho_{ED}/A^{\alpha_{e}})$ defines the lnA reconstruction line (fig. \ref{xyeu}), the linear trend in panels (c) and (f) crucially indicates that the lnA shift correlates linearly with $lg(E/A)$ across hadronic models. \\\indent
\section{Performance}
\label{performance}
For this analysis, a few selections are applied to improve the quality of the data, they are listed as below: \\\indent
(1) the number of fired ED is more than 20; \\\indent
(2) the number of charged particles detected by ED with a distance less than 200 m from the shower core after noise filtering is more than 50; \\\indent
(3) the ratio of the number of charged particles detected within a 100 m radius to that within a 40-100 m ring is larger than 2; \\\indent
(4) the reconstructed zenith angle is in the range of 0 - 40 degrees; \\\indent
(5) the reconstructed core location is inside the KM2A full-array region, and the minimum distance between the core location and the boundary of KM2A full-array is larger than 50 meter. \\\indent
This section presents the bias and resolution of the reconstructed energy and lnA for the selected sample, along with their dependence on the hadronic model. \\\indent

\subsection{Energy bias and resolution}

\begin{figure*}[htbp]
\begin{minipage}[t]{1.0\linewidth}
\includegraphics[width=0.49\linewidth]{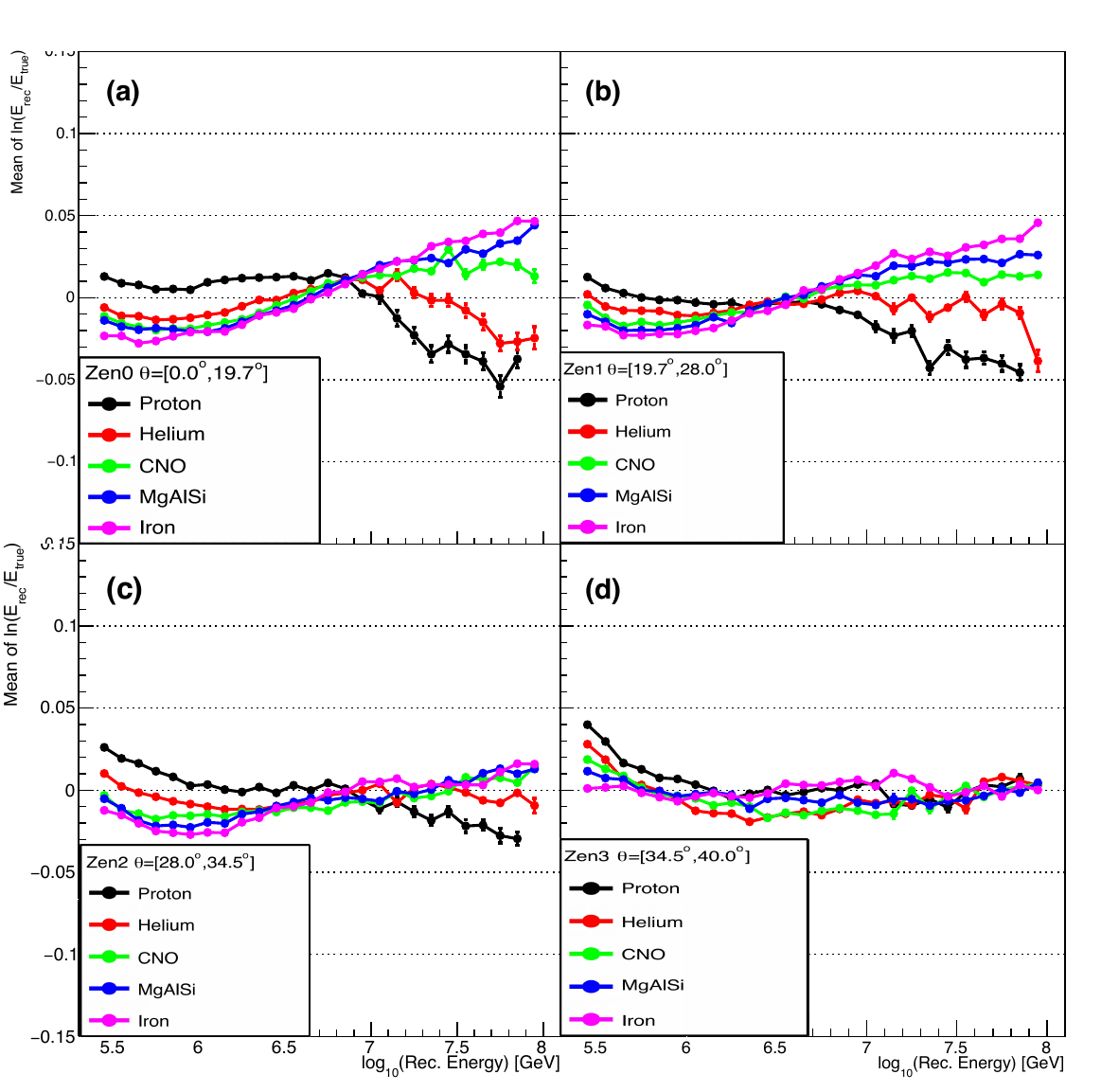}
\hspace{0cm}
\includegraphics[width=0.49\linewidth]{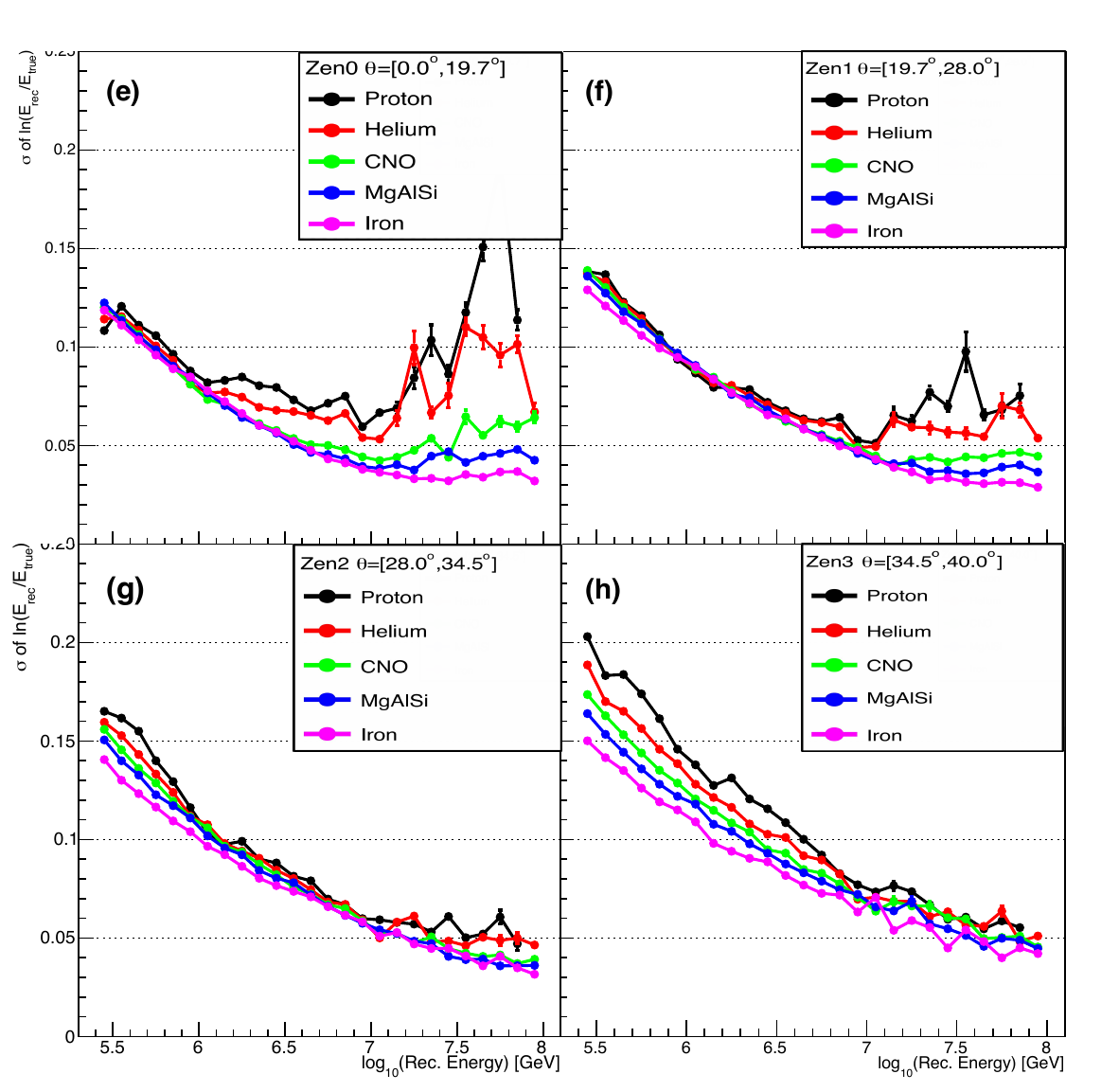}
\caption{Bias (a–d) and resolution (e–h) of reconstructed energy as a function of energy. Each panel corresponds to a different zenith angle; colors denote composition.}
\label{engrec}
\end{minipage}
\end{figure*}

The resolution and bias of the energy are defined as the mean and sigma parameter of the fitted Gaussian function to the distribution of $ln(E_{rec})-ln(E_{true})$ respectively, where $E_{rec}$ is the reconstructed energy based on the method introduced above, $E_{true}$ is the true energy in MC. The resolution and bias of the energy for samples with different type of primary particles and zenith angles are presented in fig. \ref{engrec}.\\\indent
As can be seen in fig. \ref{engrec}, The energy bias depends on the zenith angle and primary composition, while always within $\pm$5\%. The bias improves at higher zenith angles, reaching approximately $\pm$2\% for all compositions in the 35–40 degree bin. The energy resolution as a function of energy is different for different primary particles and different zenith angles. For proton nuclei, at low energy (below 10 PeV), the energy resolution of vertical direction is better than inclined direction, and the energy resolution is within 13\%. While at higher energy (above 10 PeV), the energy resolution is getting worse when the direction of primary particle becomes less inclined, especially for vertical incident particles, this is due to that LHAASO is measuring the secondaries at a fixed altitude, the best resolution is achieved around shower maximum, while above 10 PeV, LHAASO detects secondaries before shower maximum for vertically incident protons. For events with zenith angles above 28 degree, the proton energy resolution is below 10\% above 3 PeV and reaches around 5\% at the highest energy. For iron nuclei, the behavior is similar except the differences of energy resolution between different zenith angles are smaller. The energy resolution ranges from below 5\% to 15\%.
\subsection{lnA bias and resolution}
\begin{figure*}[htbp]
\begin{minipage}[t]{1.0\linewidth}
\includegraphics[width=0.49\linewidth]{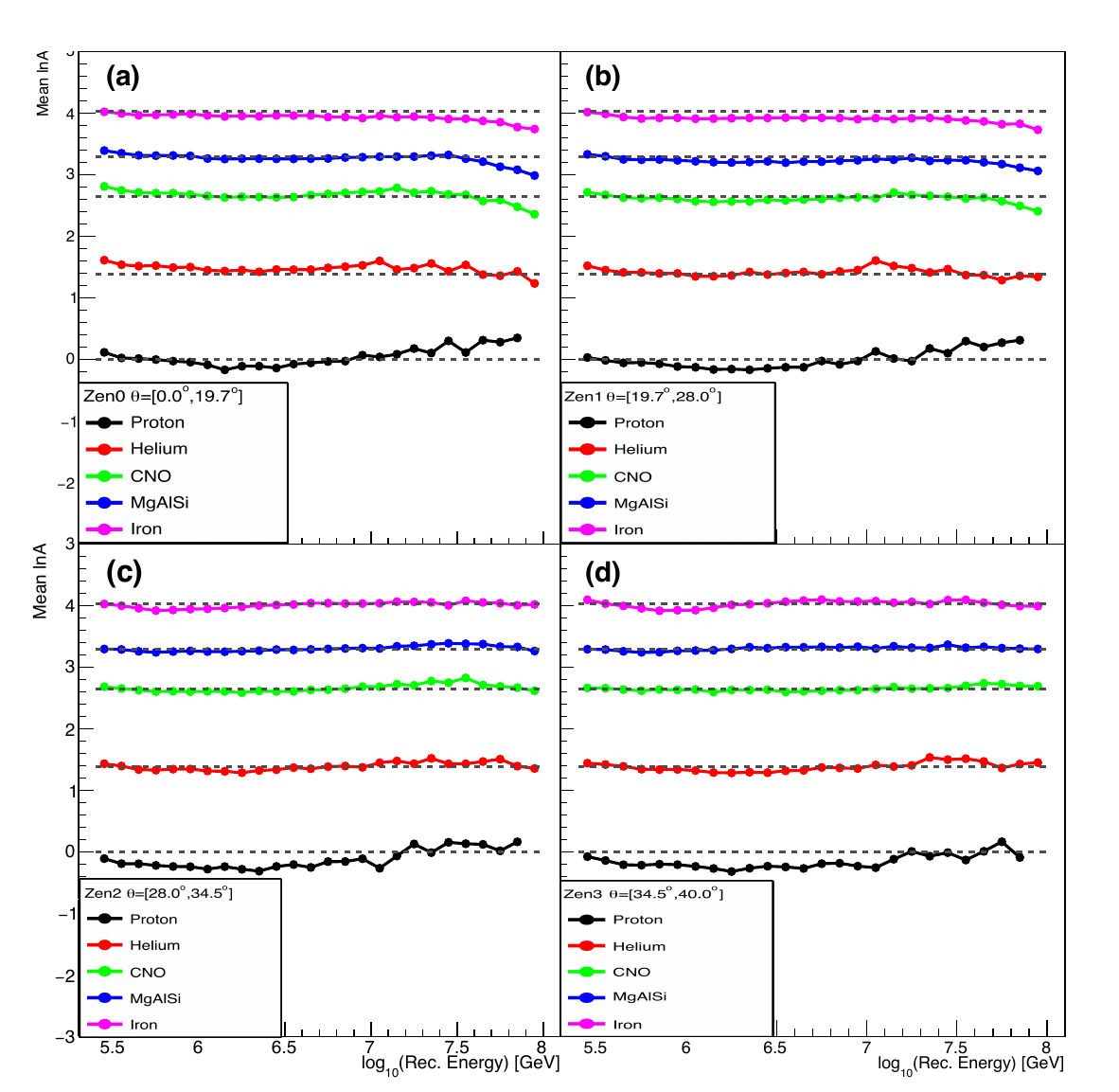}
\hspace{0cm}
\includegraphics[width=0.49\linewidth]{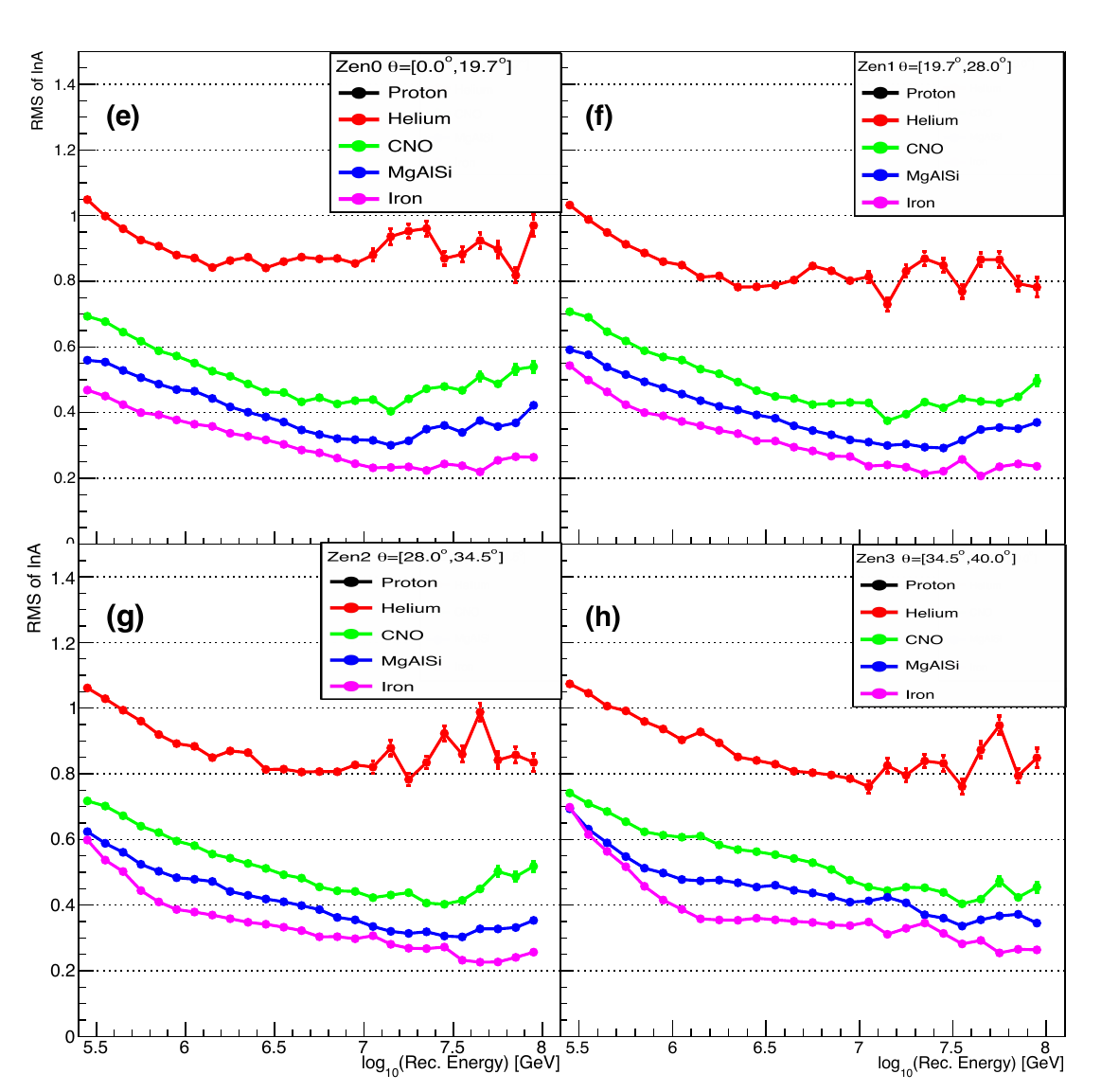}
\caption{Bias (a–d) and resolution (e–h) of reconstructed lnA as a function of energy. Each panel corresponds to a different zenith angle; colors denote composition.}
\label{lnArec}
\end{minipage}
\end{figure*}
\begin{figure*}[htbp]
\begin{minipage}[t]{1.0\linewidth}
\includegraphics[width=0.7\linewidth]{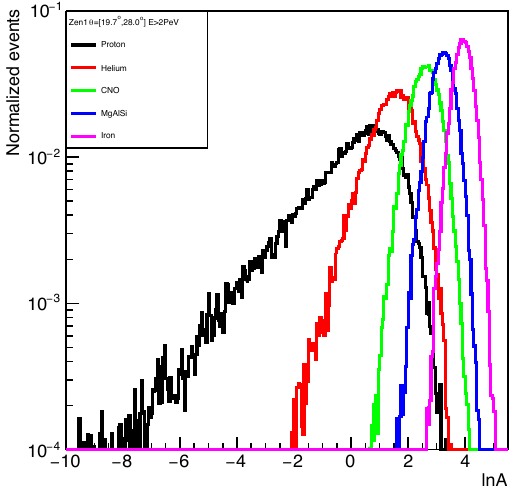}
\caption{The comparison of lnA distribution for different composition above 2 PeV.}
\label{lnAcomp}
\end{minipage}
\end{figure*}

The lnA bias and resolution are defined as the mean and standard deviation, respectively, of $lnA_{rec}-lnA_{true}$. Here, $lnA_{rec}$ is the reconstructed value from $\rho_{ED}$ and $\rho_{MD}$, and $lnA_{true}$ is the true logarithm of the mass number from the Monte Carlo simulation. These quantities are shown in fig. \ref{lnArec} for different particle types and zenith angles. \\\indent
As can be seen in fig. \ref{lnArec}, the bias of lnA is within 0.3 overall for all composition and zenith angles, For vertical incident particles, the lnA bias is within 0.2 below 10 PeV. Above 10 PeV, it starts to getting worse. While for high zenith angle bin, for example, in the range of 35-40 degree, the lnA bias is within 0.1 for all composition except proton, for proton it is within around 0.3. The bias is small compared to rms of lnA and the differences between the hadronic models shown in Fig. \ref{EAmodel}. \\\indent
The lnA resolution behaves similarly across zenith angles as a function of energy. For iron nuclei, it remains within 0.4 above PeV at all zenith angles, with the best resolution less than 0.25 (corresponding to a 25\% mass resolution) above $\sim$ 10 PeV. The lnA resolution is worse for lighter nuclei, particularly protons and helium, due to their larger shower-to-shower fluctuations. The comparison of lnA distribution for different composition is shown in Fig. \ref{lnAcomp}, as seen, the proton lnA distribution has a long left-side tail, making it distinct from other compositions. This feature can be used to select high-purity protons~\cite{lhaaso-proton}, although it results in a large RMS. \\\indent
\subsection{Density versus counts}
\begin{figure*}[htbp]
\includegraphics[width=0.8\linewidth]{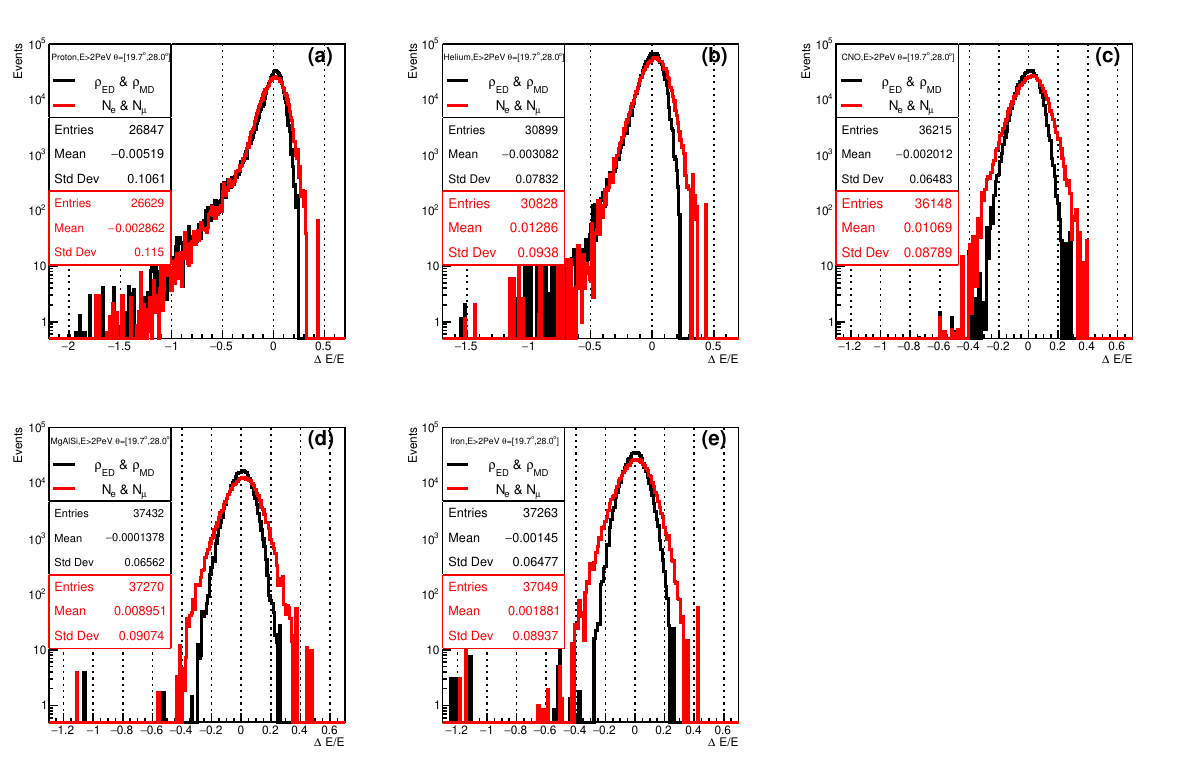}
\caption{Energy resolution functions for five mass groups above 2 PeV, comparing density-based (black) and counts-based (red) reconstruction: (a) Proton, (b) Helium, (c) CNO, (d) MgAlSi, and (e) Iron.}
\label{lnEdis}
\end{figure*}
\begin{figure*}[htbp]
\includegraphics[width=0.8\linewidth]{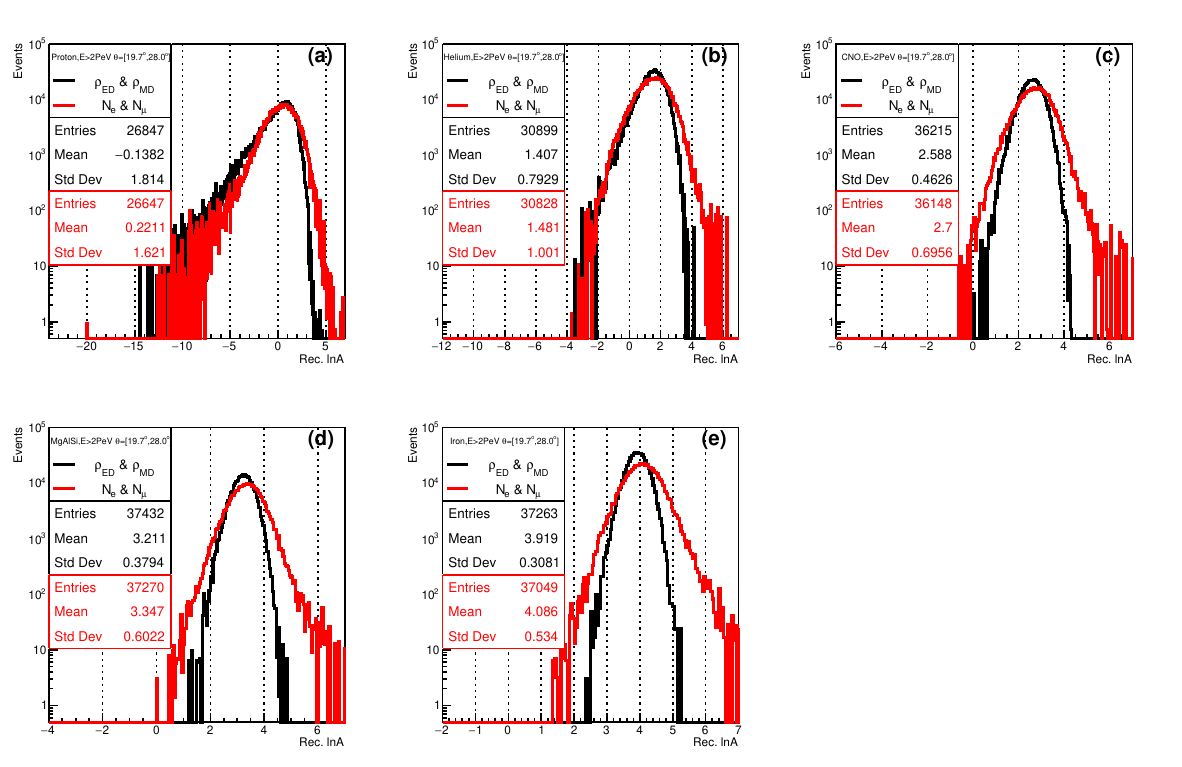}
\caption{lnA distribution for five mass groups above 2 PeV, comparing density-based (black) and counts-based (red) reconstruction: (a) Proton, (b) Helium, (c) CNO, (d) MgAlSi, and (e) Iron.}
\label{lnAdis}
\end{figure*}
\begin{figure*}[htbp]
\includegraphics[width=0.45\linewidth]{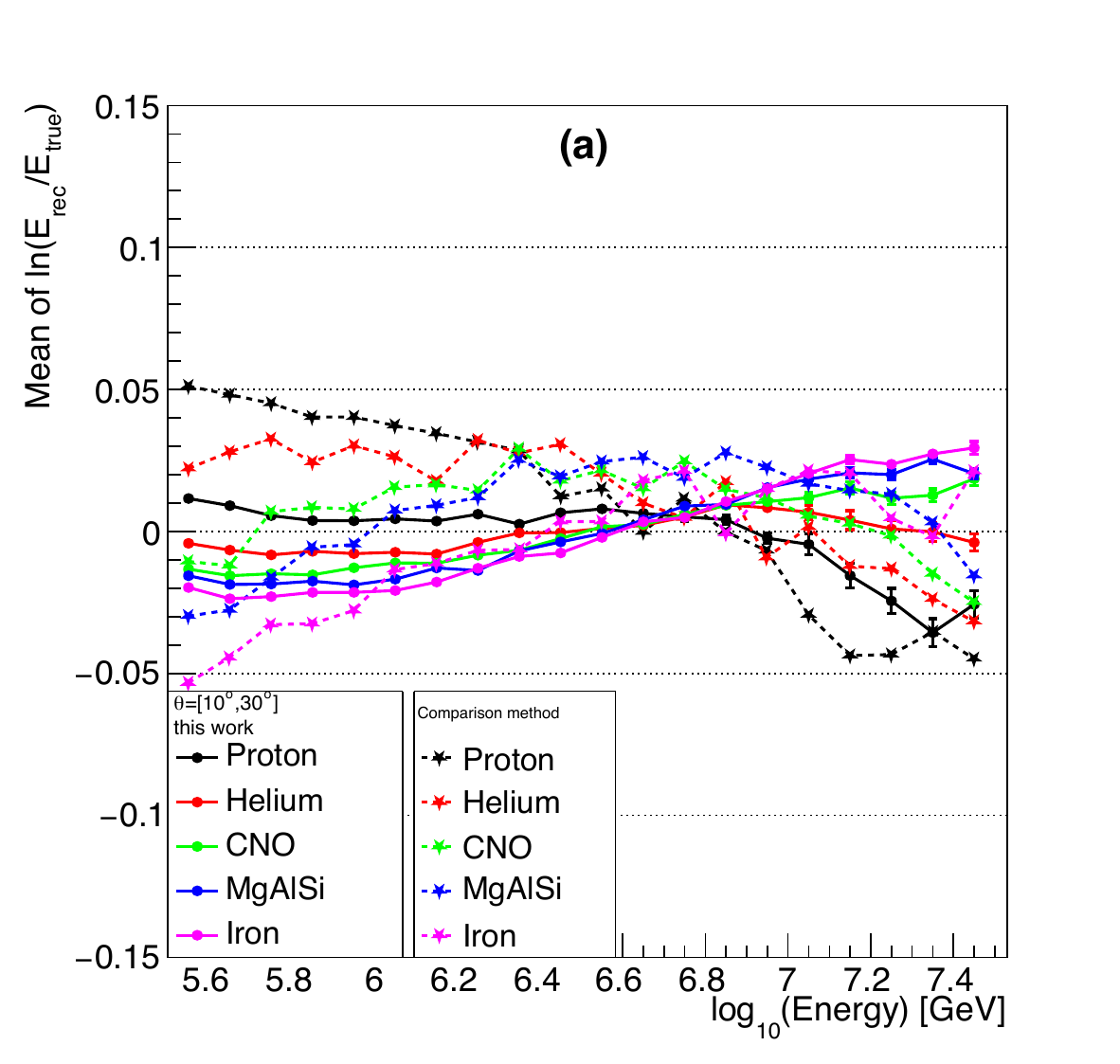}
\includegraphics[width=0.45\linewidth]{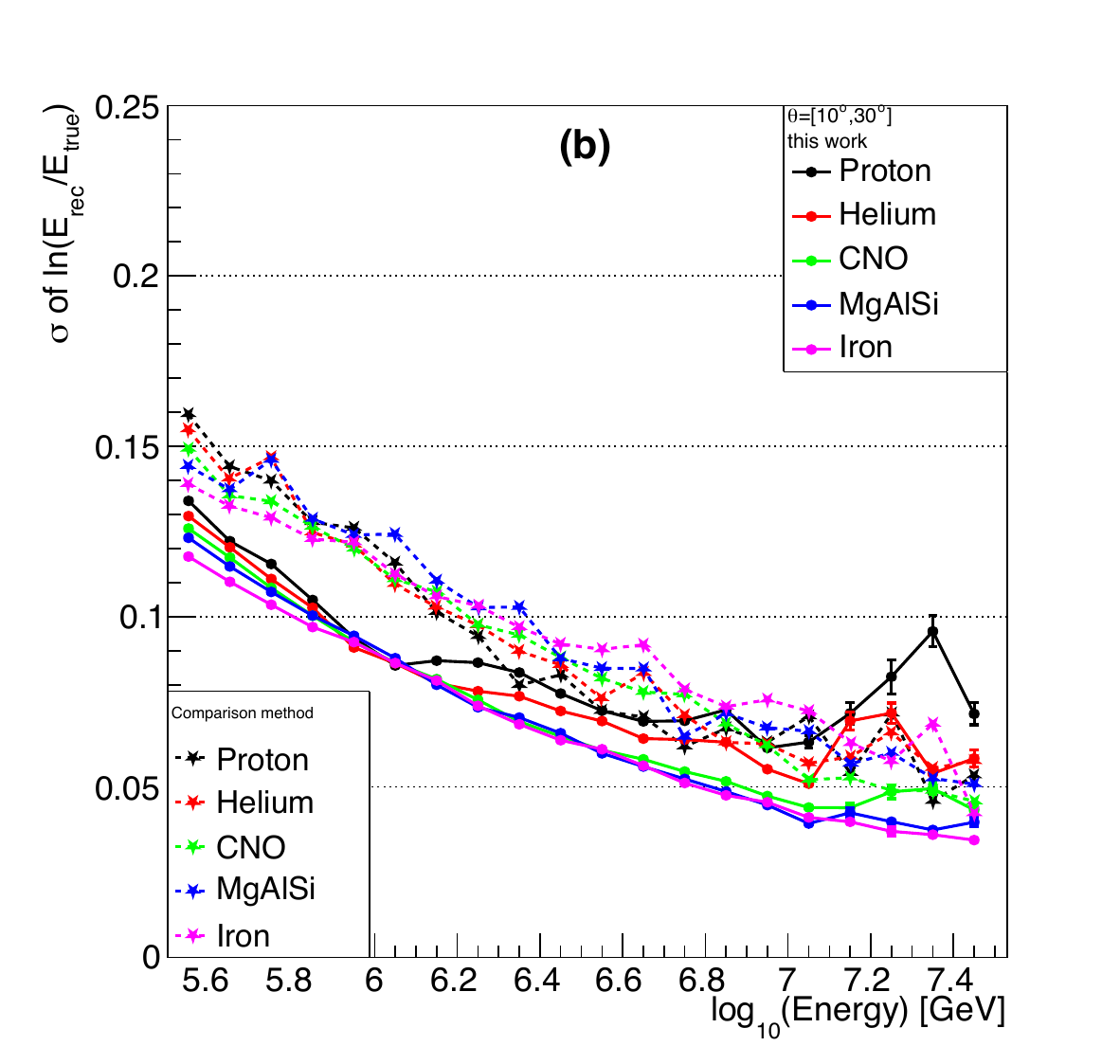}
\includegraphics[width=0.45\linewidth]{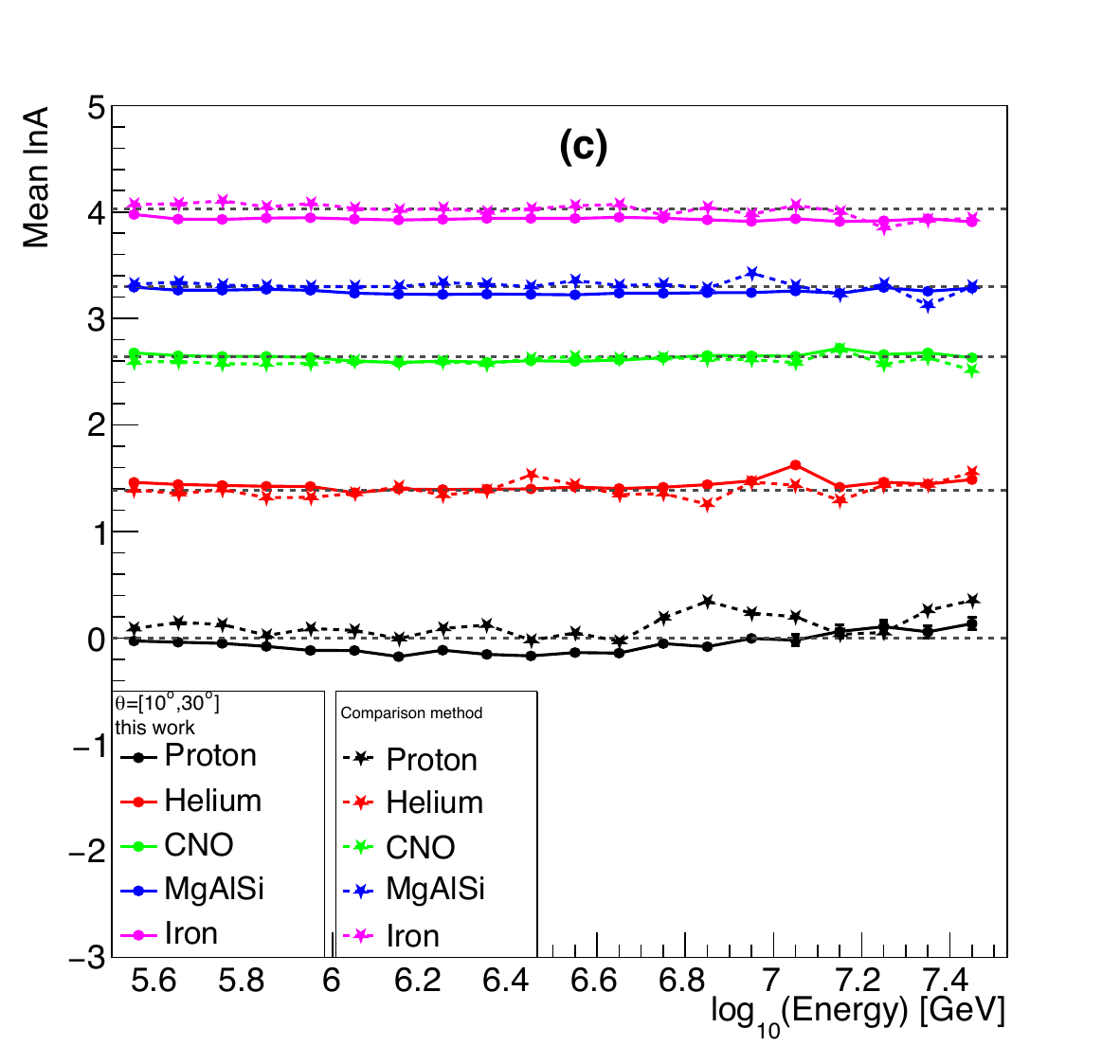}
\includegraphics[width=0.45\linewidth]{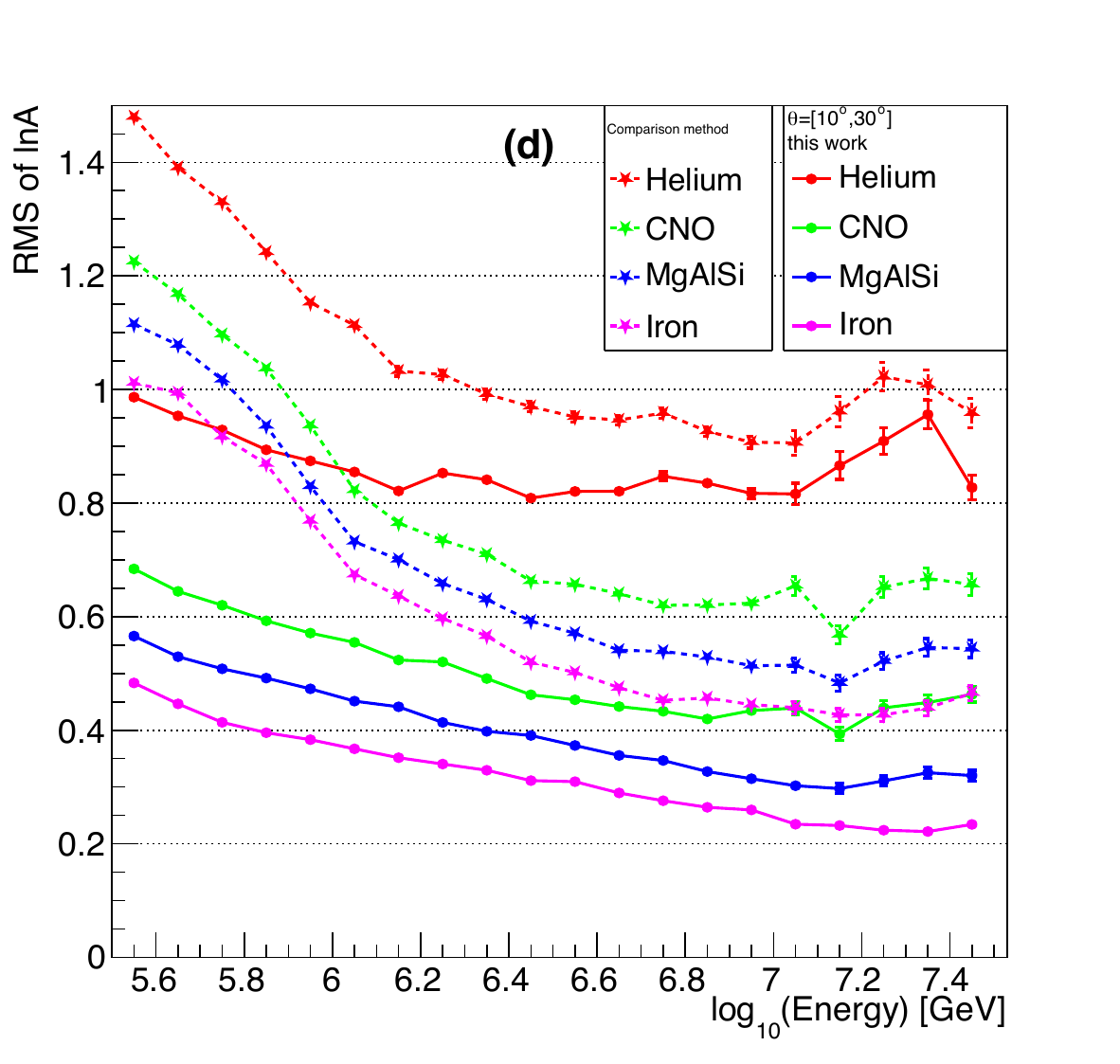}
\caption{Comparison of bias (panels a, c) and resolution (panels b, d) for reconstructed energy (panels a, b) and lnA (panels c, d), plotted as a function of reconstructed energy. Results from the present work are shown by solid dots and solid lines, while the comparison method from Ref.~\cite{lhaaso-all} is represented by stars and dashed lines. Different colors correspond to different mass compositions. The horizontal dashed line in panel (c) indicates the true lnA value for reference.}
\label{EA_compare}
\end{figure*}

The energy and lnA reconstruction is also performed using electromagnetic ($N_{e}$) and muon counts ($N_{\mu}$) within annular bands for comparison with the density-based method. Here, $N_{e}$ and $N_{\mu}$ both are the counts within 40–200 m from the shower axis. Figure \ref{lnEdis} compares the energy resolution distributions for five mass groups above 2 PeV. Figure \ref{lnAdis} compares the lnA distributions for five mass groups above 2 PeV. The density-based method improves both the energy and lnA resolution. The improvement is greater for heavy nuclei and marginal for light nuclei, with smaller differences observed when using a narrower, more centrally focused core position region.
\\\indent
To further compare against state‑of‑the‑art approaches, we adopt the existing energy and lnA reconstruction method from Ref.~\cite{lhaaso-all} — based on electromagnetic‑particle and muon counts measured by the same LHAASO‑KM2A detector — as a benchmark, which is denoted as the comparison method hereafter. Figure \ref{EA_compare} compares the energy and lnA bias and resolution across five mass groups between our density‑based method and this benchmark method. To have a fair comparison, the zenith angle range and core positions are chosen to be the same as those in Ref.~\cite{lhaaso-all}, i.e., the zenith angle ranges from 10 to 30 degrees, and core positions are restricted to inner and outer ring radii of 320 and 420 m with respect to the center of LHAASO. The data of the comparison method in panels a, b, and c of Fig.~\ref{EA_compare} are the same as those in Ref.~\cite{lhaaso-all}. The RMS of the reconstructed lnA for the comparison method in panel d of Fig.~\ref{EA_compare} is calculated based on the equation and parameters introduced in that reference. As shown, the energy and lnA biases are comparable for the two methods. Our method yields marginally better energy bias, yet slightly underestimates lnA for protons and iron nuclei. Its key advantage lies in improved energy and mass resolution, as illustrated in Fig.~\ref{EA_compare}d. This makes our method well‑suited for measuring individual energy spectra. It is also worth noting that the energy-resolution degradation for protons (or heliums) in this work starts at approximately 1 PeV; see Fig.~\ref{engrec}e and Fig.~\ref{EA_compare}b. This occurs because the proton shower maximum lies near the atmospheric overburden at the LHAASO site around 1 PeV for near-vertical incidence. At higher energies, secondaries of protons are sampled above the shower maximum, causing the energy-resolution degradation. For larger zenith angles, this effect is much weaker owing to the thicker atmospheric overburden; see Fig.~\ref{engrec}g and \ref{engrec}h. Consequently, high-energy protons and helium nuclei can be reconstructed well at larger zenith angles using this method.
\subsection{Hadronic model dependence}
\begin{figure*}[htbp]
\begin{minipage}[t]{1.0\linewidth}
\includegraphics[width=0.8\linewidth]{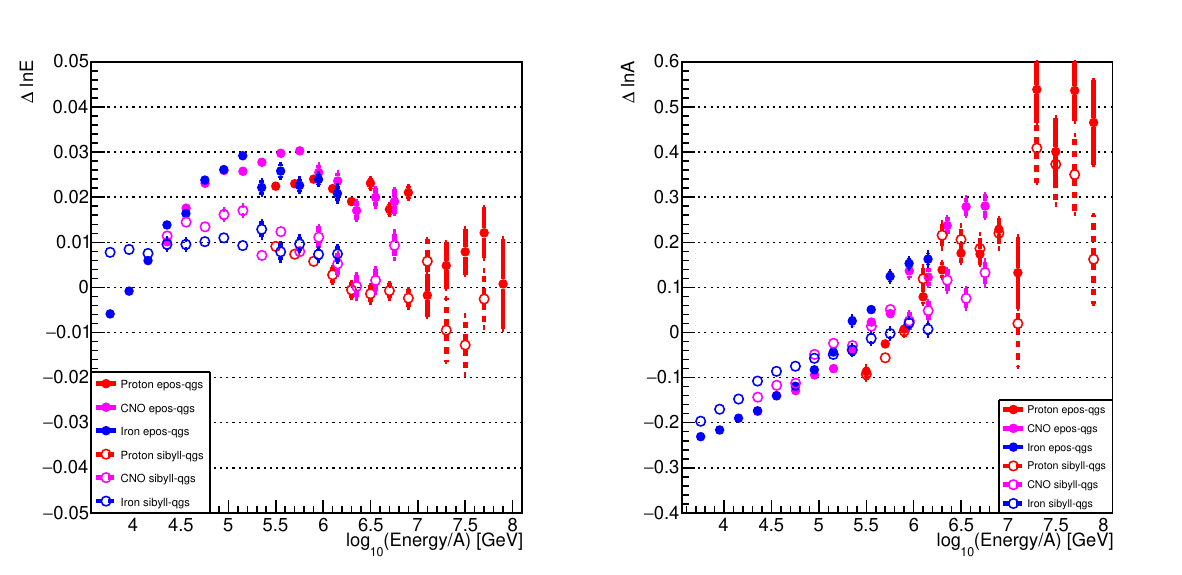}
\caption{Left panel: mean energy difference of EPOS-LHC and SIBYLL samples relative to QGSJet data in each reconstructed energy bin, versus energy/A, where A is the mass number. Different colors represent different compositions; filled and open circles correspond to EPOS-LHC-QGSJet and SIBYLL-QGSJet, respectively; Right panel: mean lnA difference of EPOS-LHC and SIBYLL samples relative to QGSJet data in each reconstructed energy bin, versus energy/A. The colors and markers are the same as in the left panel. The zenith angle is in the range $20^{\circ}$ to $28^{\circ}$. }
\label{EAmodel}
\end{minipage}
\end{figure*}
As established by experiments, extensive air shower (EAS) measurements depend on the hadronic model used in shower simulation~\cite{2008PhRvL.101q1101P, 2016PhRvL.117s2001A, 2021PhRvL.126o2002A}. Figure \ref{EAmodel} shows the mean differences in the reconstructed energy and lnA for EPOS-LHC and SIBYLL data relative to QGSJet-II-04, for zenith angles of 20 to 28 degrees. The reconstruction parameters for all models were generated from QGSJet-II-04 simulated data. \\\indent
As shown in left panel of fig. \ref{EAmodel}, the mean difference of energy of the EPOS-LHC model and SIBYLL model respect to QGSJet model are both within $\sim3\%$ in all energy range for all type of primary particles. The energy bias depends on energy per mass number (A) and is nearly independent of primary composition. \\\indent
The right panel of Fig. \ref{EAmodel} shows the mean lnA differences of the EPOS-LHC and SIBYLL models relative to QGSJet, as a function of primary energy divided by A. The differences scale approximately linearly with $lg(E/A)$ and are nearly independent of primary composition. These behaviors arise primarily because secondary properties scale with $\lg(E/A)$, as predicted by the superposition model (Eqs. \ref{funED} and \ref{funMD}).\\\indent
For iron nuclei, the lnA difference can reach $\sim$0.2, comparable to the lnA resolution for iron, which could influence the observed lnA distribution. Consequently, comparing the lnA distributions in the heavy nuclei region in cosmic-ray data to those from various hadronic models may provides an effective test of the models. \\\indent

\section{Summary}
\label{summary}
Events detected by LHAASO-KM2A full array are reconstructed by fitting the lateral distribution of electromagnetic particles and muons derived from the ED hits and MD hits. The density of electromagnetic particles at 100 meter away from the shower axis and the density of muons at 150 meter away from the shower axis from the fitted lateral distribution were used to reconstruct the energy and lnA, compared to other variables that used the particle number in an annular band, those two have smaller shower-to-shower fluctuation. \\\indent
Using the verified superposition model, we developed an independent energy and lnA reconstruction method for LHAASO-KM2A. They are reconstructed using two universal, composition- and energy-independent calibration lines. The dependence on these parameters is the main source of systematic uncertainty. Its performance is summarized below. \\\indent
For energy reconstruction, the bias is within $\pm5\%$ for all masses and zenith angles from 300 TeV to 100 PeV. A minimum bias of $\pm 2\%$ occurs at zenith angles of $35^{o}-40^{o}$. Above 1 PeV, the energy resolution ranges from below 5\% to $\sim 15\%$. The bias from hadronic model dependence is within $\sim 3\%$.  \\\indent
For lnA reconstruction, the bias is within 0.3 for all masses and zenith angles from 300 TeV to 100 PeV. The resolution of lnA for iron is less than 0.4 above 1 PeV at all zenith angles, with the best resolution less than 0.25 (corresponding to a 25\% mass resolution) above $\sim$ 10 PeV. Which is similar to the lnA difference between hadronic models, this could influence the observed lnA distribution. Comparing the lnA distributions in heavy nuclei region in cosmic-ray data with those from various models provides an effective test of hadronic models. The lnA resolution degrades for lighter nuclei. Although the proton lnA distribution has a large RMS, its distinct long left-side tail enables the selection of high-purity protons. \\\indent
The hadronic model dependencies of energy and lnA are also studied in this work. The dependencies scale with $lg(E/A)$ and are nearly independent of primary composition, as predicted by the superposition model. \\\indent
In summary, this method yields energy and lnA biases comparable to those of the existing method, and achieves good energy and lnA resolution, which provides important input for measuring individual energy spectra. \\\indent
\begin{acknowledgments}
The authors would like to thank all staff members who work at the LHAASO site above 4400 meters above sea level year-round to maintain the detector and keep the electrical power supply and other components of the experiment operating smoothly. We thank the Chengdu Management Committee of Tianfu New Area for their constant financial support of research with LHAASO data. We are especially grateful to Prof. L.L. Ma and colleagues from the LHAASO Collaboration for their valuable suggestions and comments. This work is supported by the National Natural Science Foundation of China (Grant number 12205244).
\end{acknowledgments}

\bibliographystyle{apsrev4-2}
\bibliography{main}
\end{document}